\documentclass[11pt]{article}
\usepackage[margin=1in]{geometry}
\usepackage{amsmath}
\usepackage{amssymb}
\usepackage{amsthm}
\usepackage{dcolumn}
\usepackage{mdwlist}
\usepackage{bm}
\usepackage{graphicx}
\usepackage{cite}
\usepackage{subfig}
\usepackage{calc}
\usepackage{hyperref}
\usepackage[capitalise]{cleveref}

\newcommand{\Vector}[1]{\bm{\mathrm{#1}}}
\newcommand{\Matrix}[1]{\bm{\mathrm{#1}}}
\DeclareMathOperator{\sgn}{\mathrm{sgn}}
\DeclareMathOperator{\Adj}{\mathbf{Adj}}
\DeclareMathOperator{\Tr}{\mathrm{Tr}}

\begin{document}
\title{Generalized Hamiltonian Dynamics and Chaos in Evolutionary Games on Networks}

\author{Christopher Griffin
\footnote{
	Applied Research Laboratory,
    The Pennsylvania State University,
    University Park, PA 16802
    }
\and Justin Semonsen
\footnote{
	Department of Mathematics,
	Rutgers University,
    Piscataway, NJ 08854
    }
\and  
Andrew Belmonte
\footnote{
	Department of Mathematics / The Huck Institutes for the Life Sciences,
	The Pennsylvania State University,
	University Park, PA 16802
}
}

\date{\today}

\maketitle

\begin{abstract} We study the network replicator equation and characterize its fixed points on arbitrary graph structures for $2 \times 2$ symmetric games. We show a relationship between the asymptotic behavior of the network replicator and the existence of an independent vertex set in the graph and also show that complex behavior cannot emerge in $2 \times 2$ games. This links a property of the dynamical system with a combinatorial graph property. We contrast this by showing that ordinary rock-paper-scissors (RPS) exhibits chaos on the 3-cycle and that on general graphs with $\geq 3$ vertices the network replicator with RPS is a generalized Hamiltonian system. This stands in stark contrast to the established fact that RPS does not exhibit chaos in the standard replicator dynamics or the bimatrix replicator dynamics, which is equivalent to the network replicator on a graph with one edge and two vertices ($K_2$). 
\end{abstract}



\section{Introduction}

The Hamiltonian approach to the dynamics of complicated, interacting systems has had substantial success in providing a mathematical understanding of the world, yielding key results in classical, celestial, statistical, and quantum mechanics. Surprisingly, the evolutionary dynamics of games, which have been studied extensively over the last 40 years (see e.g.~\cite{TJ78,Ze80,SS83,NM92,Wei95,HS98,HS03,NS04,EGB16,FS16}) can in some cases be shown to possess a Hamiltonian structure, stemming from the dynamical description implicit in the replicator equation for the evolution of strategy choices \cite{EA83,Hof96,SAF02}. The replicator equation is one of several differential equations proposed for evolutionary games \cite{HS98,sand2010,EGB16} and it has also been generalized to many situations, including the coevolutionary dynamics of multiple games (different payoff matrices) \cite{EA83,Hof96,TCH05,AD14,PG16}.  The simplest case is the bimatrix formulation \cite{EA83,Hof96}, with dynamics given by:
\begin{equation}
\left\{
\begin{aligned}
\dot{x}_i &= x_i\left(\Vector{e}_i - \Vector{x}\right)\mathbf{A}\Vector{y}\\
\dot{y}_i &= y_i\left(\Vector{e}_i - \Vector{y}\right)\mathbf{B}\Vector{x}.
\end{aligned}
\right.
\label{eqn:Bimatrix}
\end{equation}
Here we have two interacting species, with strategy proportion vectors $\Vector{x}(t)$ and 
$\Vector{y}(t)$ and corresponding game matrices $\Matrix{A}$ and $\Matrix{B}$ (fully generalized in \cite{PG16}). 
For the bimatrix replicator, it has been shown that all interior equilibria (corresponding to coexisting strategies or phenotypes within a species) are unstable \cite{EA83}. 
As we discuss below, when each species plays the same game ($\mathbf{A} = \mathbf{B}^T$), this is identical to the network replicator (\cref{eqn:NetReplicator}) on the graph with two vertices and one edge (the graph $K_2$).

When $\mathbf{A} = \mathbf{B}$ is symmetric, and the bimatrix equation is further symmetrized to a single species with $\Vector{x} = \Vector{y}$, then
\cref{eqn:Bimatrix} becomes the ordinary replicator equation. Zeeman and others have shown that chaotic behavior does not occur in the ordinary replicator  with three or fewer strategies \cite{Ze80,SS83}; however chaotic behavior can emerge with four strategies \cite{S92}. 
Thinking of \cref{eqn:Bimatrix} as network replicator on $K_2$ \cite{ON06,SF07,QSGU18}, Sato and others have shown \cite{SAF02,SC03,SAC05} that chaotic behavior can emerge in three strategy games, however not the \textit{ordinary} rock-paper-scissors (RPS) game \cite{SAF02}. 
Similarly, for the classical replicator, work by \cite{EA83,Hof96,SAF02,FS16} makes it clear that ordinary RPS and its generalizations are Hamiltonian systems, but do not exhibit chaos.  

\begin{figure}
\centering
\includegraphics[width=0.45\columnwidth]{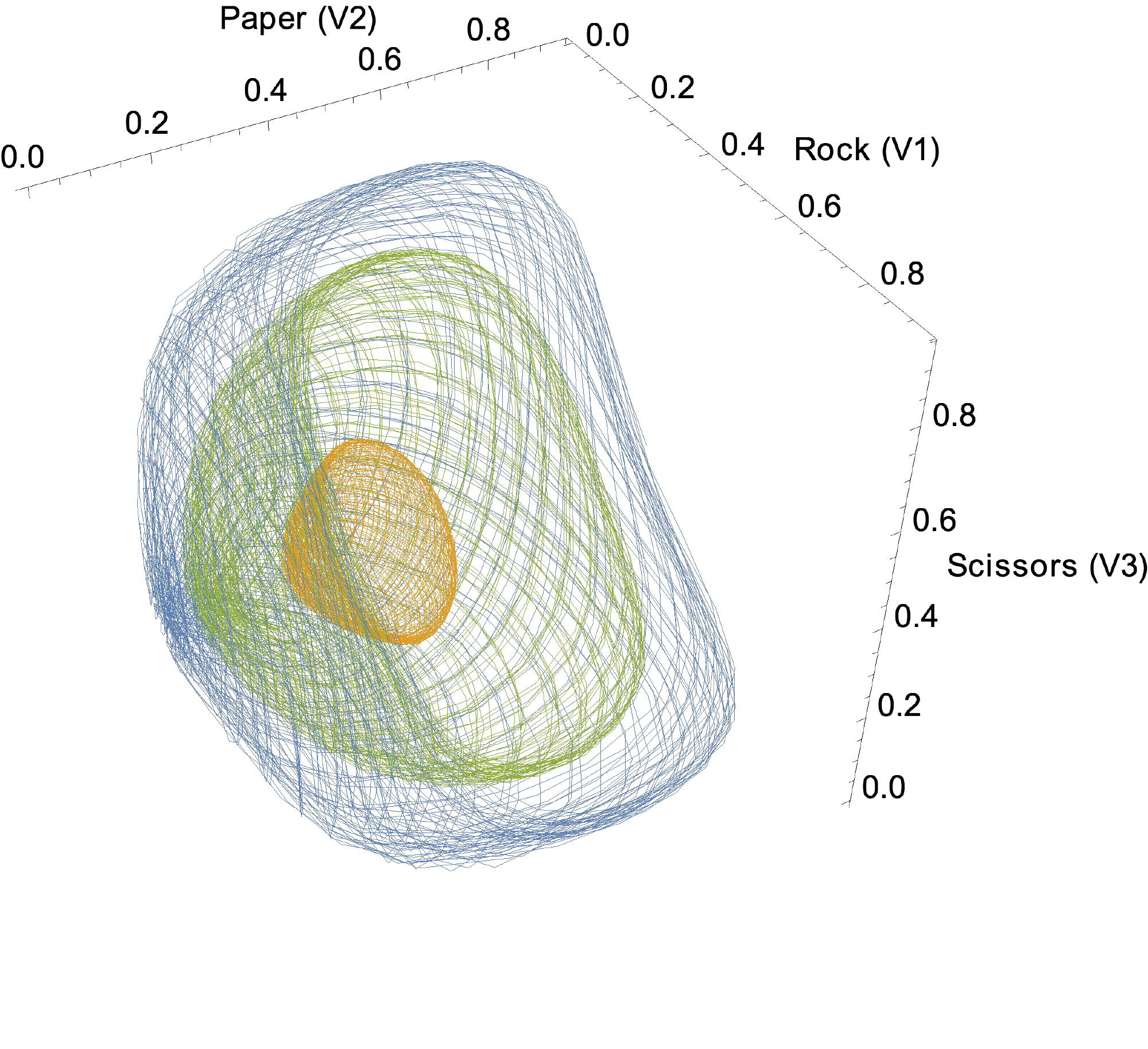}
\includegraphics[width=0.45\columnwidth]{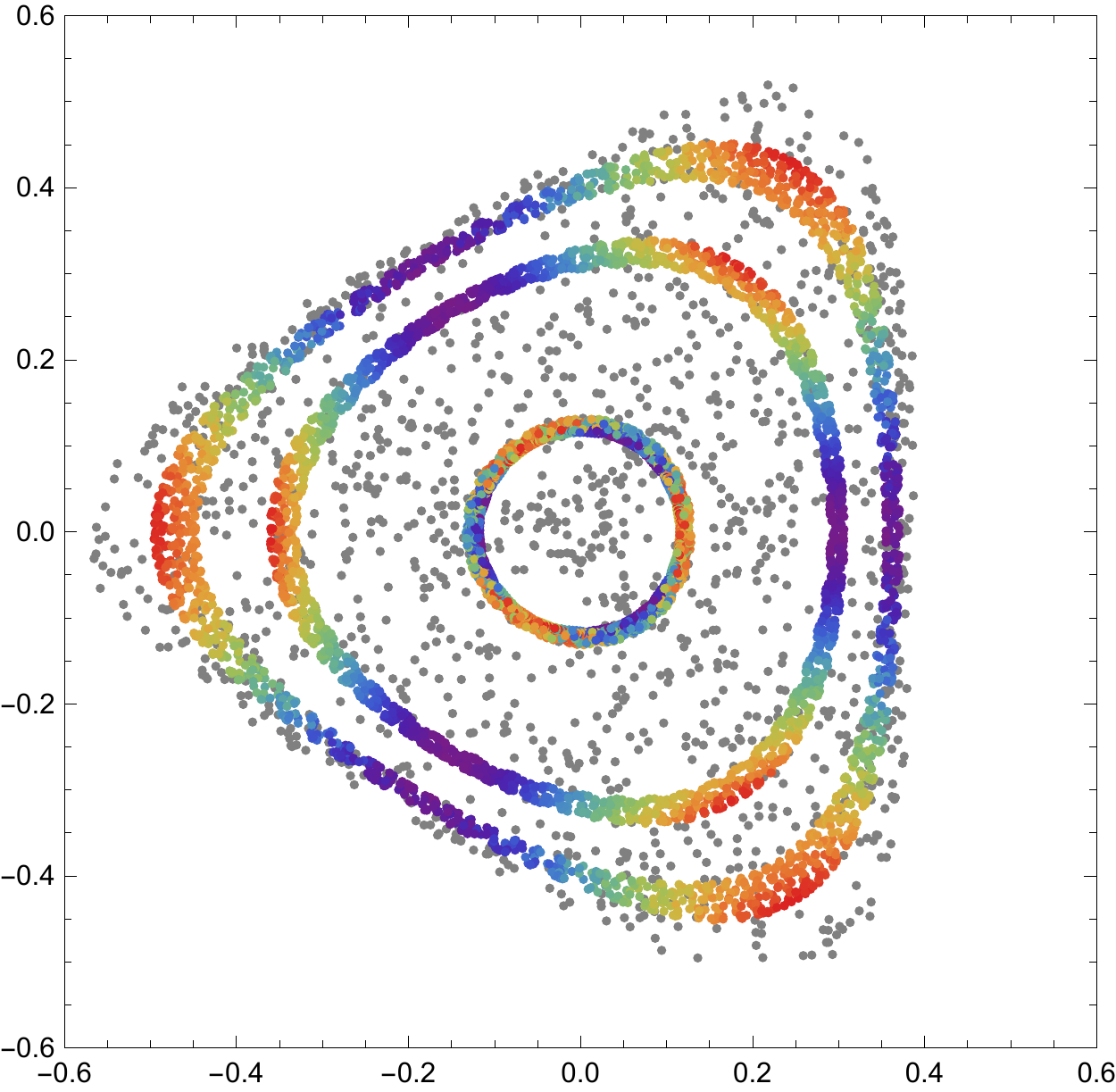}
\caption{Rock-Paper-Scissors dynamics in the network replicator: 
(L) projection of 9-dimensional trajectories following (nested) surfaces in $\mathbb{R}^3$; (R) Poincar\'e sections of four different initial conditions, showing three quasiperiodic (color spectrum) and one chaotic (grey).}
\label{fig:Fig1}
\end{figure}

In this paper, we show that by enlarging the network from two nodes (the bimatrix case, $K_2$) to three (the three-cycle, $K_3$), the network replicator admits chaotic behavior for ordinary RPS, as illustrated in \cref{fig:Fig1}. Moreover, the network replicator equation possesses a \textit{generalized Hamiltonian structure} on an arbitrary graph. Our approach also leads to a surprising link between a combinatorial aspect of the graph structure and the asymptotic behavior of the time-evolving strategy for two strategy games. To our knowledge, this is the first connection made between the asymptotic or chaotic behavior in an evolutionary game or coevolutionary dynamics on graphs and the graph structure itself.

Let $G = (V,E)$ be a graph consisting of $n > 1$ vertices, each representing a player, with $V = \{1,\dots,n\}$. For simplicity, we will use a common (single) payoff matrix $\Matrix{A} \in \mathbb{R}^{m \times m}$. Following \cite{QSGU18}, each vertex is a player who may use a mixed strategy of dimension $m$ in a symmetric game (repeatedly) played against neighboring vertices. In this case, the network replicator equation is
\begin{equation}
\dot{x}_{ij} =  x_{ij}\left(\sum_{k \in N(i)}\left(\Vector{e}_j - \Vector{x}_i \right) \cdot \Matrix{A}\Vector{x}_k\right)
\label{eqn:NetReplicator}
\end{equation} 
where $N(i)$ are the neighbors of vertex $i$. 
The network replicator has been studied recently in the control literature \cite{MMQ14,OPQ14,MM15,RC17,BOQ17} with a special focus on $2 \times 2$ games. 
While there are several derivations of the network replicator in the literature, we provide a straightforward derivation from a population model perspective in \cref{sec:AppendixA}.

\section{Asymptotic Behavior of $2 \times 2$ Games}
We first show that the dynamics of $2 \times 2$ games are in some sense simple and related to certain combinatorial properties of the graph structure. We begin by characterizing the fixed points of \cref{eqn:NetReplicator} in $2 \times 2$ games.  Without loss of generality (see \cref{sec:AppendixB}), assume the payoff matrix is of the form:
\begin{displaymath}
\Matrix{A} = \begin{bmatrix} 0 & r\\s & 0\end{bmatrix}.
\end{displaymath}
This includes anti-coordination games ($r >0, s >0$) and Prisoner's Dilemma-type games 
($rs < 0$).
The fact that we have only two strategies ($j \in \{1, 2\}$) simplifies the analysis substantially. Let $x_i \in [0,1]$ be the fraction of the time player $i$ plays Strategy 1, and $\Vector{x}_i = \langle{x_i, 1-x_i}\rangle$. Then the network replicator for node $i$ becomes
\begin{equation}
\dot{x_i} =
x_i(1-x_i)\left(\sum_{j\in N(i)} r-(r+s)x_j\right).
\label{eqn:2stratDynamics}
\end{equation}
Thus any fixed point $\Vector{x}^* = \langle{x_1^*,\dots,x_n^*}\rangle$ of 
\cref{eqn:2stratDynamics} must have, for each vertex $i$, either $x_i = 0$, $x_i = 1$ 
(the pure strategies), or 
\begin{equation}
\frac{1}{|N(i)|}\sum_{j \in N(i)} x_j = \frac{r}{r+s}
\label{eqn:MixedFixedPoint2}
\end{equation}
(assuming $r+s \neq 0$). This final condition specifies the average of the neighboring strategies surrounding $i$.

The stability of any $\Vector{x}^*$ is determined by the eigenvalues of the corresponding Jacobian matrix. For the network replicator, these eigenvalues must be real for $2 \times 2$ games  (see \cref{sec:AppendixB}). To examine the stability of $\Vector{x}^*$, we define $S \subset V$ to be the set of vertices for which the player is playing a mixed strategy, i.e. if $i \in V$, then $x_i^* \in (0,1)$. Let $G[S]$ denote the subgraph of $G$ generated by the vertices in $S$.  We now analyze the fixed points in two distinct cases: 

(i) {\it $r$ and $s$ have opposite signs (Prisoner's Dilemma type).}
The right hand side of \cref{eqn:MixedFixedPoint2} cannot be in $[0,1]$, 
so there can be no vertices with a mixed strategy. Thus $S = \emptyset$, and 
$\Vector{x}^*$ is a pure strategy fixed point.
In this case the Jacobian matrix 
is diagonal, consequently $\Vector{x}^*$ is hyperbolic and admits no circulation. 
Moreover, the defect strategy will be asymptotically stable for all players.

\begin{figure}[!t] 
\centering
\includegraphics[width=0.55\columnwidth]{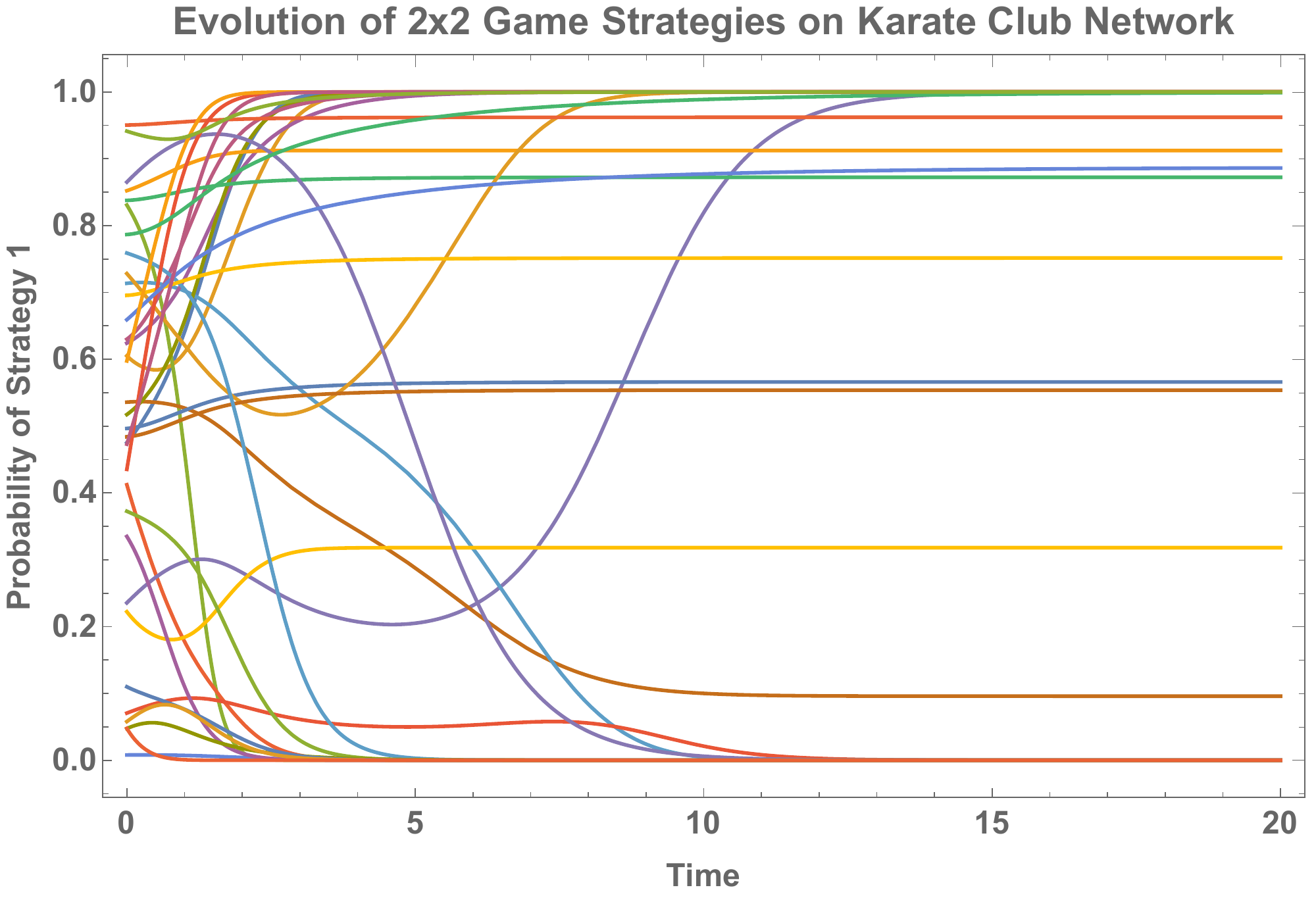}
\includegraphics[width=0.4\columnwidth]{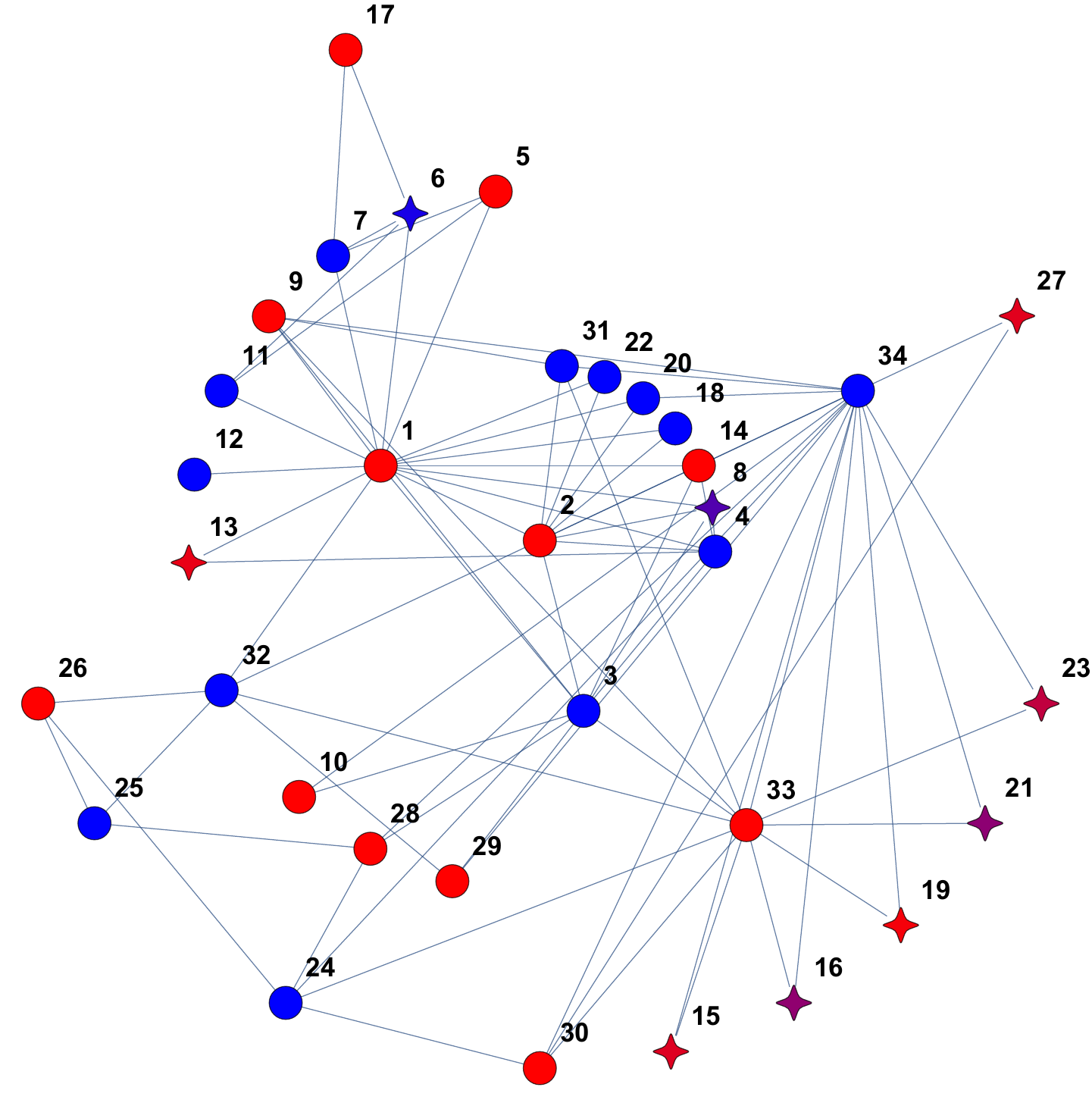}
\caption{The evolution of the network replicator on the Karate Club graph for a 
$2\times 2$ anti-coordination game: 
(Left) convergence to a steady state $\Vector{x}^*$ from random initial conditions;
(Right) structure of $\Vector{x}^*$, showing pure (circles) and mixed (star-shape) strategy vertices, with color distinguishing strategy choice (see text).}
\label{fig:2x2KarateClub}
\end{figure}

(ii) {\it $r$ and $s$ have the same sign.}
Assume $r,s > 0$ without loss of generality; it is now possible for $S$ to be non-empty. Analysis of the Jacobian matrix shows that $\Vector{x}^*$ is unstable whenever $G[S]$ has an edge (see \cref{sec:AppendixB}). Thus the asymptotic dynamics is linked to a combinatorial property of the graph structure: the existence of an independent set of vertices in $G$. This is illustrated in \cref{fig:2x2KarateClub}.

Thus neither circulation nor chaotic behavior is possible in the $2 \times 2$ payoff matrix case. In general, solutions always converge to a (neutrally) stable fixed point as a result of the compactness of the manifold $\Delta_k^m$, since the eigenvalues of the Jacobian are always real. 
Furthermore, any vertices that play mixed strategies must form an independent set in the graph (i.e. are not connected by an edge); this effectively
limits the number of vertices that can play a mixed strategy at equilibrium.  
Thus the network replicator predicts that, in any ecological network defined by a $2 \times 2$ game, no two interacting species can both include coexisting strategies at equilibrium.

\section{Chaotic Dynamics in Simple 3-Strategy Games}
While the dynamics of the network replicator are simple for $2 \times 2$ games, we show  numerically that chaotic behavior emerges in the classic (symmetric) RPS game when played on a three node network with three edges (the 3-cycle $K_3$); note that no  chaotic behavior is observed on the two node network, as shown (unintentionally) in \cite{SAF02} for the classic RPS as a bimatrix game. 
In the standard replicator, this game has a single interior elliptic fixed point. Generalizations of the RPS game are discussed in \cite{HS03}, whose dynamics are entirely classified by Zeeman, who showed that no limit cycles can emerge \cite{Ze80}.

Consider the network replicator on $K_3$, with the three nodes playing RPS defined by the payoff matrix 
\begin{equation}
\mathbf{A} = \begin{bmatrix}0 & -1 & 1 \\ 1 & 0 & -1\\ -1 & 1 & 0\end{bmatrix}
\label{eqn:RPSA}
\end{equation}
Straightforward analysis shows that the system has an infinite number of fixed points (see \cref{sec:AppendixC}), which can be classified into three pure strategies, a continuum of boundary strategies (where one strategy is chosen with zero probability), and one interior fixed point $\left\langle{\tfrac{1}{3},\tfrac{1}{3},\tfrac{1}{3}}\right\rangle$. 

Define the vector valued function $\Vector{F}:\Delta_3^3 \to \mathbb{R}^9$ so that the network replicator dynamics are $\dot{\Vector{x}} = \Vector{F}(\Vector{x})$. Simple computation shows that shows that $\nabla \cdot \Vector{F} = 0$, i.e., the system is conservative. As a consequence, the interior fixed point must be a non-linear (elliptical) center, and the boundary fixed points are non-attracting. We show that this property of the network replicator leads to periodic, quasi-periodic and chaotic dynamics, a result similar to what is found in \cite{SAC05}, but with simpler dynamics. We also note, this is a variation on the result in \cite{EA83} which argues that the ordinary replicator on two species preserves a certain volume form. 

\begin{figure}[!b] 
\centering
\includegraphics[width=0.48\columnwidth]{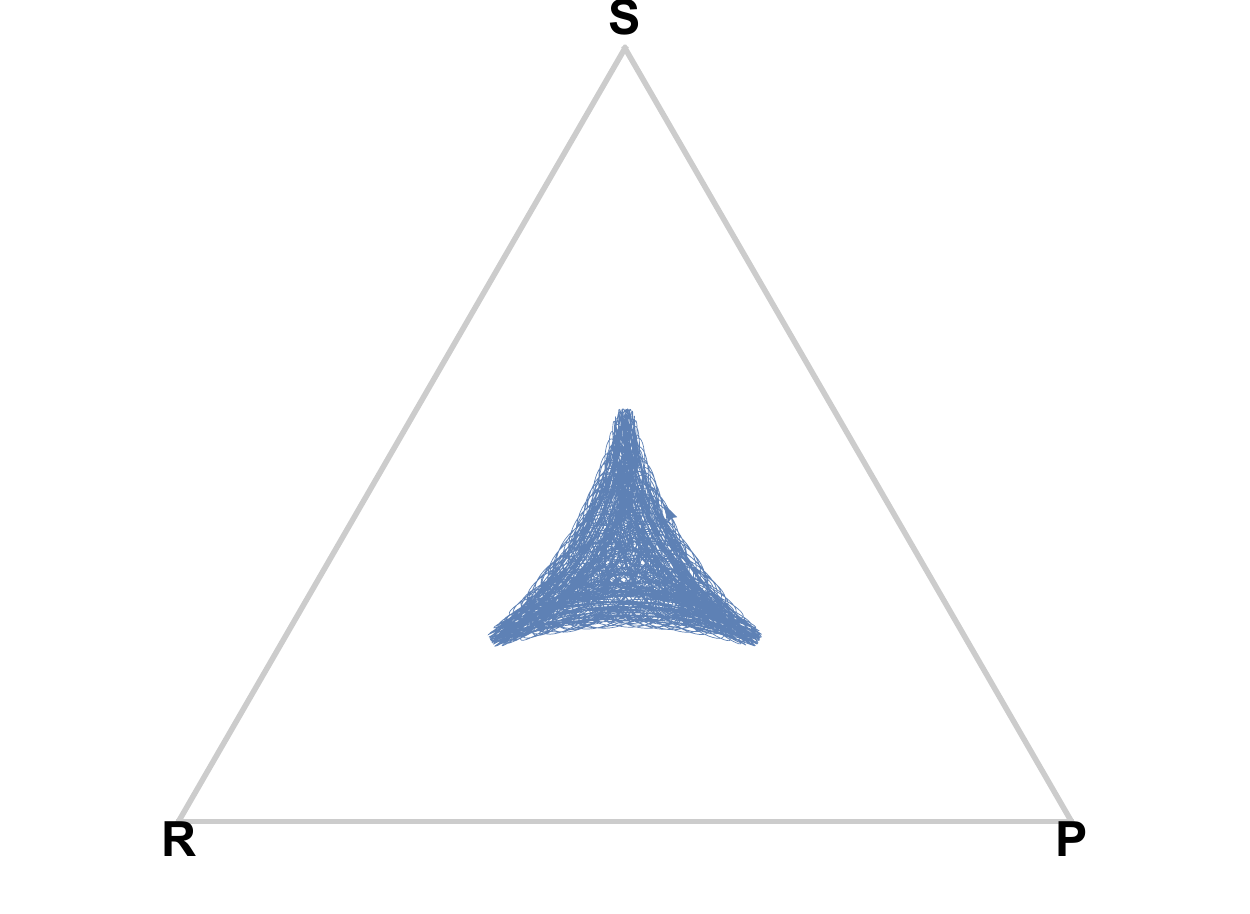}
\includegraphics[width=0.48\columnwidth]{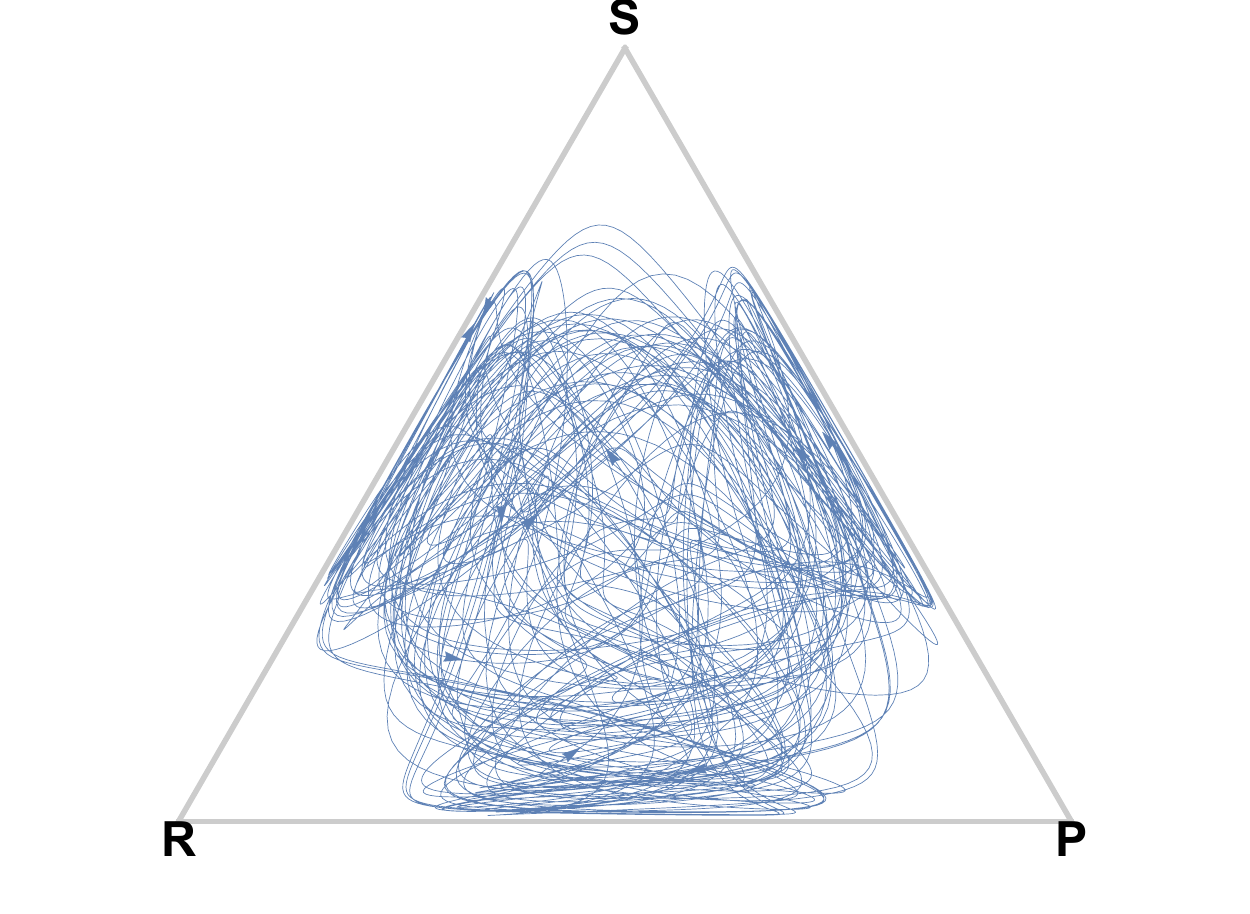}
\caption{Rock-paper-scissors on a $K_3$ network: phase portraits for Vertex 1 with 
 initial conditions close to (Left) and far from (Right) the interior fixed point.
}
\label{fig:PhasePortraits1}
\end{figure}

Long phase portraits from various starting points illustrate both quasi-periodic and chaotic motion. The phase portraits in \cref{fig:PhasePortraits1} were constructed using a ternary transform on the dynamics of Vertex $1$ alone for the $K_3$ network, and shows that chaotic behavior seems to emerge as the initial condition is moved further from the interior fixed point. 
A corresponding three-dimensional trajectory slice is shown \cref{fig:Fig1} (Left).
These surfaces are symmetric and illustrate the relationships between rock at Vertex 1, paper at Vertex 2 and scissors at Vertex 3. A two dimensional Poincar\'{e} section is shown in \cref{fig:Fig1} (Right) with the corresponding chaotic trajectory shown in \cref{fig:PhasePortraits1}, in which densely packed orbits appear relatively well-behaved when the initial condition is close to the interior fixed point. However, when the orbit is started further away, it oscillates filling up more space. This is qualitatively similar to the double pendulum, which when started close to its hanging equilibrium displays simple motion, but exhibits chaotic motion when released far away from the equilibrium point \cite{S14}. Simple, neutrally stable orbits also exist, as we show in the \cref{sec:AppendixD}.

\begin{figure}[!t] 
\centering
\includegraphics[width=0.8\columnwidth]{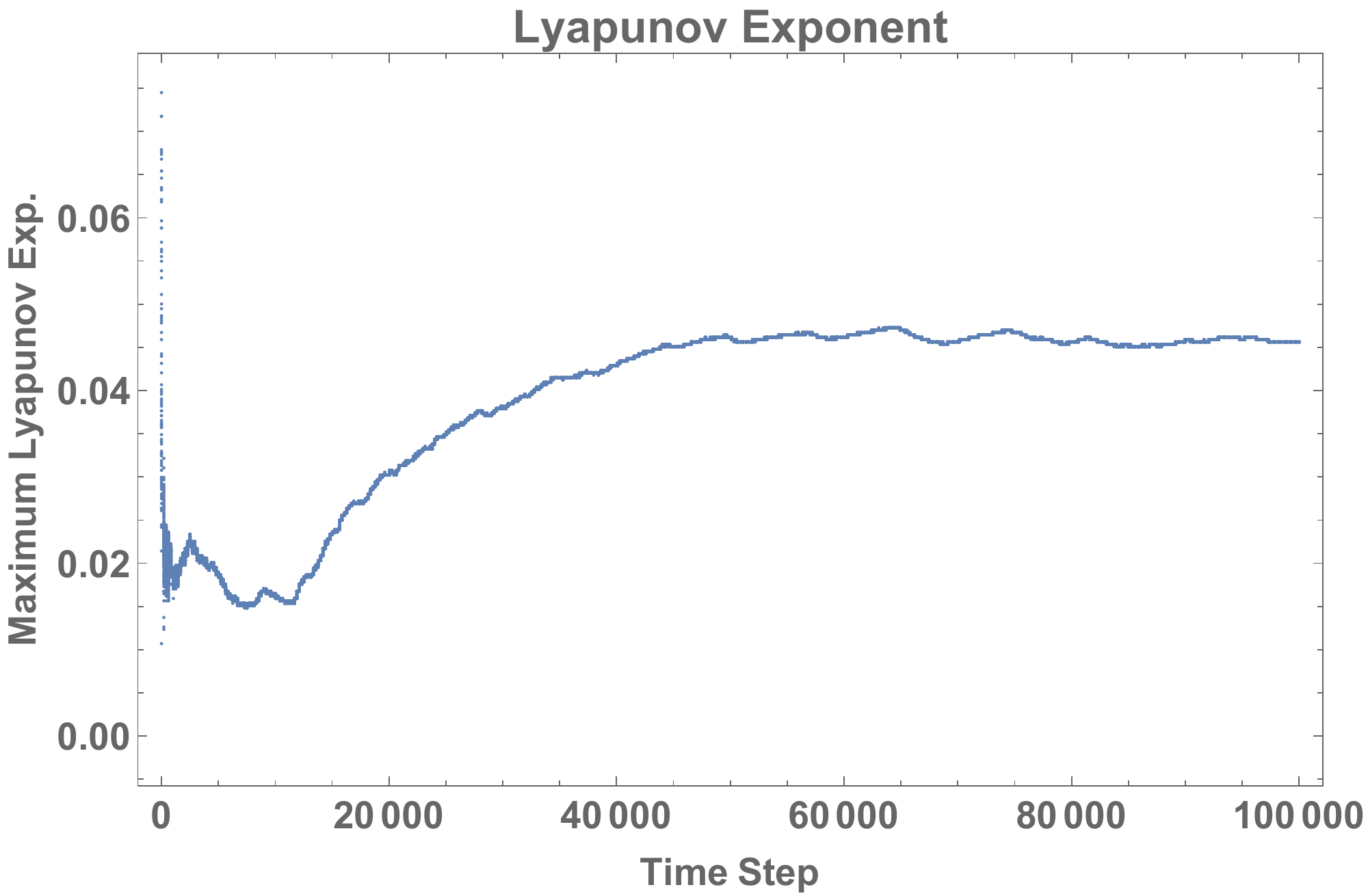}\\
\includegraphics[width=0.9\columnwidth]{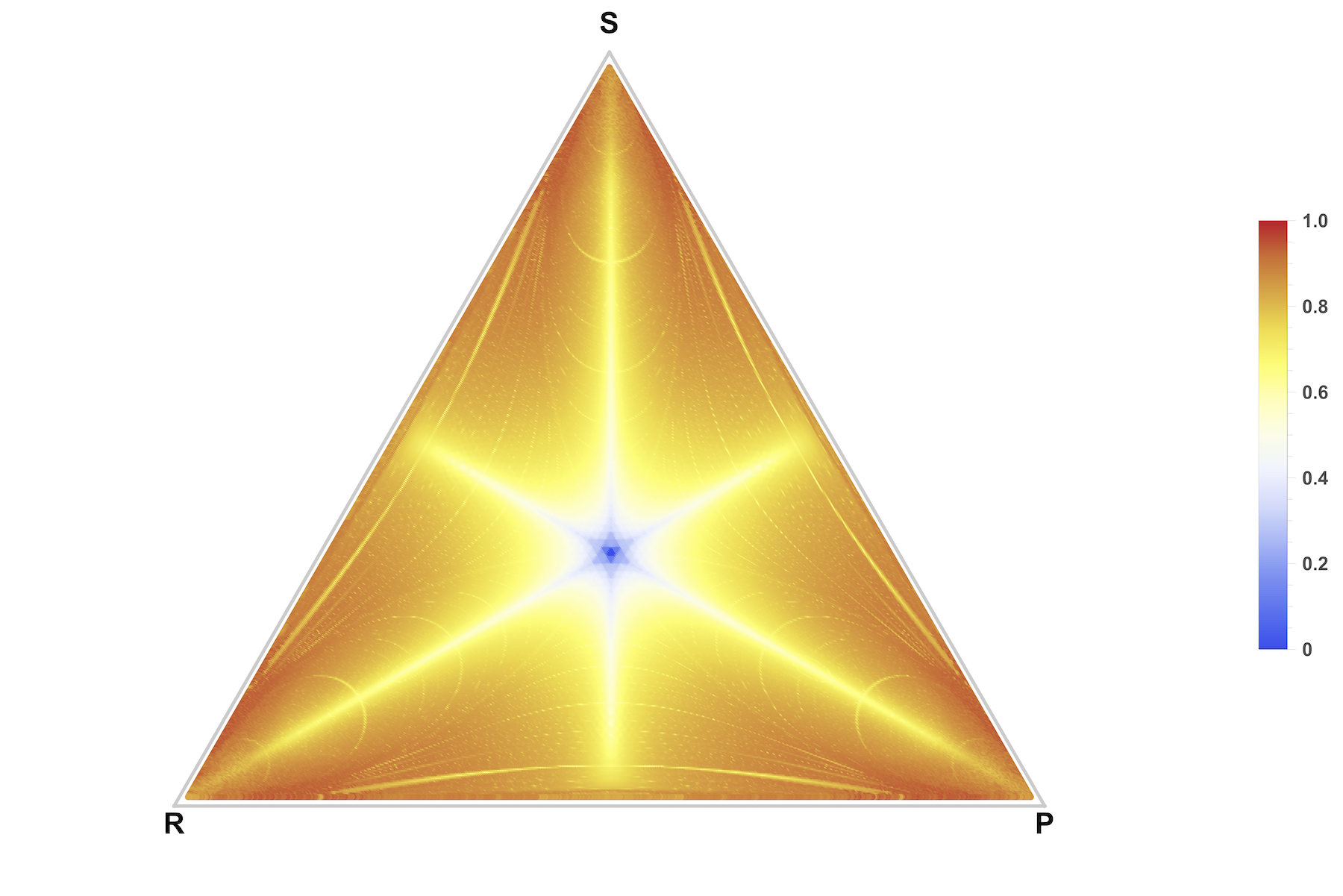}
\caption{Sensitive dependence on initial conditions in the RPS game: (Top) computed maximum Lyapunov exponent for the dynamics \cite{WSSV85};
(Bottom) density plot showing (discrete) entropy of trajectories as a function of initial point in the simplex.}
\label{fig:ChaoticBehavior}
\end{figure}

To quantify (and in some sense prove numerically) that this system is chaotic, we computed the Lyapunov exponents using the technique in \cite{WSSV85} and implemented in \cite{S96,BZ08,K18}. The maximum Lyapunov exponent in this case is shown in \cref{fig:ChaoticBehavior} (Top). The fact that the maximum Lyapunov exponent is positive and the domain of the dynamics is compact (i.e., $\Delta_3^3$) is sufficient to show that the system exhibits chaos \cite{Shaw81,WSSV85}. The sum of the computed Lyapunov exponents is $1.3\times 10^{-7}$, consistent with the conservative nature of the flow in phase space (see e.g., Page 57 of \cite{So95}). We illustrate the sensitive dependence on initial conditions in \cref{fig:ChaoticBehavior} (Bottom) by computing the (discrete) entropy of trajectories with various initial conditions. The figure displays the fine structure associated with chaotic behavior. (Details are provided in the \cref{sec:AppendixE,sec:AppendixF}.)

\section{Generalized Hamiltonian Dynamics}
Motivated by the presence of quasi-periodic orbits and the emergence of chaotic behavior in this system, we show that a generalized Hamiltonian exists for a diffeomorphic transformation of RPS on a general graph. To help explain the complex conjugate momenta identified in the generalized Hamiltonian, we show that the linearized behavior of the RPS game on $K_3$ near the interior fixed point is a degenerate Hamiltonian system with more readily explainable (ordinary) conjugate momenta.

We first consider the general case of an arbitrary graph $G = (V,E)$ with $n$ vertices and using an arbitrary payoff matrix $\Matrix{A}$. We derive a generalized Hamiltonian dynamics for a diffeomorphic transformation of the network replicator (see \cref{sec:AppendixG} for details). Applying the approach in \cite{Hof96,SAC05,SC03}, where each strategy proportion is normalized by the last nonzero strategy, we define:
\begin{equation}
u_{i,j} = \log\left(\frac{x_{i,j}}{x_{i,n}}\right).
\end{equation}
Following Crutchfield \cite{SAC05,SC03}, this can be interpreted in an information-theoretic way, since each $x_{i,j}/x_{i,n}$ is just a relative probability; i.e each log-probability is an information measure for each vertex $i$. Moreover, this is a diffeomorphism on the interior of the phase space $\Delta_m^n$. Using this transformation, the modified dynamics are:
\begin{equation}
\dot{u}_{i,j} = \sum_{k \in N(i)} \frac{\left(\mathbf{e}_j - \mathbf{e}_m\right)\cdot\mathbf{A}\exp(\mathbf{u}_k)}{1+\sum_{l\neq m} \exp(u_{k,l})}.
\label{eqn:InteriorU}
\end{equation}
Using the RPS payoff matrix and simplifying yields:
\begin{equation}
\left\{
\begin{aligned}
\dot{u}_{i,1} &= \sum_{k \in N(i)} \left(1 - \frac{3e^{u_{k,2}}}{1+\sum_{j\neq m} e^{u_{k,j}} }\right) \\
\dot{u}_{i,2} &= \sum_{k\in N(i)} \left(-1 + \frac{3e^{u_{k,1}}}{1+\sum_{j\neq m} e^{u_{k,j}}} \right).
\end{aligned}
\right.
\label{eqn:GenHamSys}
\end{equation}

\begin{figure}
\centering
\includegraphics[width=0.8\columnwidth]{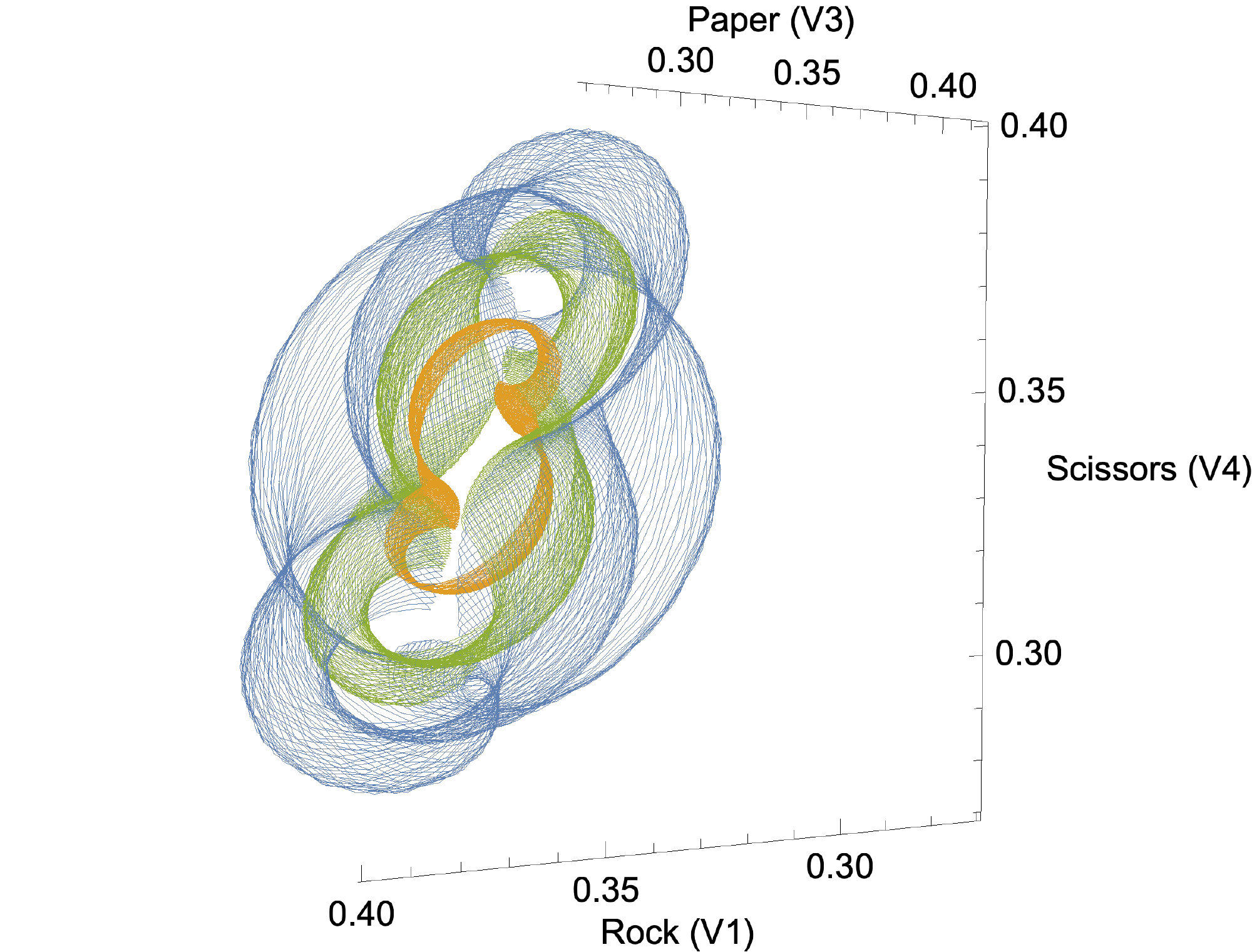}\\	
\vspace{2mm}
\includegraphics[width=0.8\columnwidth]{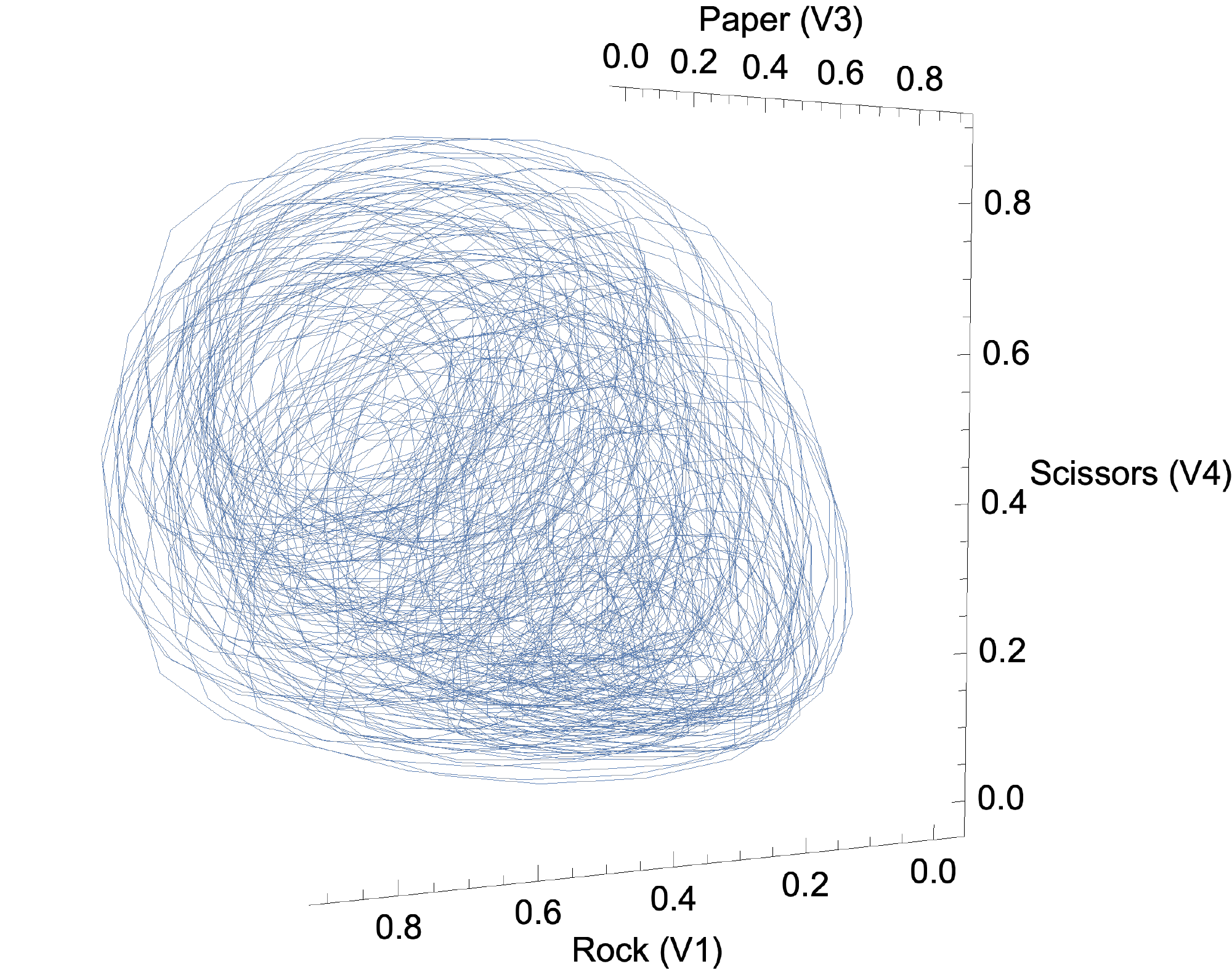}
\caption{Network replicator phase portraits for RPS on the four-cycle graph $K_4$:
(top) trajectories for initial conditions near the interior fixed point;
(bottom) chaotic trajectories for initial conditions further away.}
\label{fig:PhasePortraits2}
\end{figure}

Examining \cref{eqn:GenHamSys}, we see that $u_{i,1}$ is in a sense conjugate to a nonlinear combination of $u_{k,2}$ ($k \in N(i)$) while $u_{i,2}$ is similarly conjugate to a nonlinear combination of $u_{k,1}$. This is made explicit by defining:
\begin{equation}
\mathcal{H} = \sum_{i}\sum_{j} u_{i,j} - \sum_{i} 3\log\left(1 + e^{u_{i,1}} + e^{u_{i,2}} \right)
\end{equation}
Differentiating this generalized Hamiltonian shows that:
\begin{equation}
\text{for } i = 1,2 \left\{
\begin{aligned}
\dot{u}_{i,1} &= \sum_{k \in N(i)} \frac{\partial\mathcal{H}}{\partial u_{k,2}}\\
\dot{u}_{i,2} &= \sum_{k \in N(i)} -\frac{\partial\mathcal{H}}{\partial u_{k,1}}
\end{aligned}
\right.
\label{eqn:GeneralizedHamiltonian}
\end{equation}
The existence of a generalized Hamiltonian explains the presence of chaotic behavior far from the interior elliptic fixed point (\cref{fig:ChaoticBehavior}), and also indicates that for RPS, the network replicator provides an example of a generalized Hamiltonian system satisfying Lioville's Theorem. \cref{fig:PhasePortraits2} shows this generalized Hamiltonian chaos in the complete four species network $K_4$. As in $K_3$, the trajectories are well behaved when the initial conditions are near the interior fixed point (\cref{fig:PhasePortraits2}-top), but chaos seems to emerge for initial condition further away (\cref{fig:PhasePortraits2}-bottom).

This relationship between coordinates and conjugate momenta can be better understood conceptually by linearizing the network replicator around the RPS interior fixed point, which leads to a degenerate Hamiltonian system. We illustrate this for $K_3$ here, but the approach is similar for arbitrary graphs. Let $x_{i,3} = 1 - x_{i,1} - x_{i,2}$ for $i=1,2,3$. 
This reduces the dimension of the network replicator to six. Linearizing this system near the elliptic interior fixed point yields:
\begin{align*}
\dot{x}_{i,1} &= \frac{1}{3}\left(\sum_{j \neq i} -x_{j,1} - 2 x_{j,2}\right)\\
\dot{x}_{i,2} &= \frac{1}{3}\left(\sum_{j \neq i} 2 x_{j,1} + x_{j,2}\right)
\end{align*} 
If we define the following (conjugate momenta) variables:
\vspace{-3mm}
\begin{displaymath}
p_{i,1} =  \sum_{j \neq i} x_{j,2}
\qquad
p_{i,2} = \sum_{j \neq i} x_{j,1}
\end{displaymath}
then the linearized reduced dimensional system can be written as:
\begin{equation}
\forall i\left\{
\begin{aligned}
3\dot{x}_{i,1} &= -p_{i,2} - 2p_{i,1}\\
3\dot{x}_{i,2} &= 2p_{i,2} + p_{i,1}\\
3\dot{p}_{i,1} &= 4x_{i,1} + 2x_{i,2} + \sum_{j \neq i} 2x_{j,1} + \sum_{j \neq i} x_{j,2} \\
3\dot{p}_{i,2} &= -2x_{i,1} - 4x_{i,2} - \sum_{j \neq i} x_{j,1} - \sum_{j \neq i} 2x_{j,2} 
\end{aligned}\right.
\label{eqn:linsys}
\end{equation}
The conjugate momenta for $x_{i,1}$ have game-theoretic meaning: the $p_{i,1}$  are the strategies of other players that result in non-zero payoff for player 1, while
the $p_{i,2}$ are strategies resulting in non-zero payoffs for player 2.

The Hamiltonian for this linearized system is:
\begin{multline}
3\mathcal{H}_0 = 
\sum_{i} p_{i,1}^2 + p_{i,2}^2 + p_{i,1}p_{i,2} + \sum_{i} 2x_{i,1}^2 + 2x_{i,2}^2 +\\
\sum_{j} 2x_{i,1}x_{i,2} + \sum_{i}\sum_{j > i} 2x_{i,1}x_{j,1}  + \\
\sum_{i}\sum_{j > i}2x_{i,2}x_{j,2} + \sum_{i}x_{i,2}\sum_{j\neq i} x_{j,1}.
\end{multline}
Thus the reduced dimensional system behaves like a degenerate (12 dimensional) Hamiltonian system near the fixed point. Consequently, we expect to see quasi-periodic orbits tracing foliated $n$-tori reasonably close to the interior fixed point (\cref{fig:Fig1}). A similar analysis shows that on $K_4$, near the interior fixed point, the linearized system exhibits degenerate 16 dimensional Hamiltonian dynamics (see \cref{sec:AppendixH}). This result should generalize to arbitrary graph structures.

\section{Conclusions}
What we have seen here is that there is a fundamental difference between two-strategy and three-strategy games in network replicator dynamics.
%
%
In the two strategy case the dynamics are simple: there is no circulation in phase space, and trajectories correspondingly must always converge to some stable fixed point. The stability of these underlying fixed points is related both to the payoff matrix and the structure of an independent set composed of vertices playing mixed strategies. Our results raise an interesting question on the relationship between the combinatorial properties of graphs and equilibria of the network replicator, since determining the independence number of a graph is NP-hard \cite{Karp72}. In contrast, chaotic behavior emerges in ordinary rock, paper scissors when played on the 3-cycle. For any graph with more than two vertices, the network replicator is a generalized Hamiltonian system for the RPS game. We hypothesize that the resulting nested manifolds observed near the interior fixed point are generalized KAM surfaces. To support this, we show for $K_3$ that the linearized dynamics near the fixed point results in a degenerate Hamiltonian system in 12 dimensional space.

While the well-known KAM Theorem applies most directly to systems with a proper non-integrable Hamiltonian, there may be extensions of the KAM theorem for more generalized Hamiltonian dynamics, such as the type we have found here. Beyond this, there remains the question of whether there is any deeper meaning to the Hamiltonian structure of these equations that might involve the information or entropy of evolving strategy choices in network evolutionary systems \cite{SC03,SAC05}.

\section*{Acknowledgement} JS was supported in part by a Penn State CSRE Research Grant. CG and AB were supported in part by the National Science Foundation under grant CMMI-1932991. AB acknowledges the hospitality of the Weizmann Institute of Science, Dept of Physics of Complex Systems. 

\appendix

\section{Derivation of the Network Replicator}\label{sec:AppendixA}
We use $\mathbf{x} = \langle{x_1,\dots,x_n}\rangle$ to denote a column vector in $\mathbb{R}^n$. Let $\Delta_{m}$ denote the $m-1$-dimensional simplex embedded in $\mathbb{R}^{m}$ defined by:
\begin{equation}
\Delta_n = \left\{\mathbf{x} \in \mathbb{R}^{m} : \sum_{i=1}^{m} x_i = 1 \text{ and } 0 \leq x_i \leq 1\right\}.
\end{equation}

Let $G = (V,E)$ be a graph consisting of $n > 1$ vertices. For simplicity, let $V = \{1,\dots,n\}$. Following \cite{QSGU18}, each vertex is a player (type) who may use a mixed strategy in a symmetric game (repeatedly) played against other vertices and governed by the payoff matrix $\Matrix{A} \in \mathbb{R}^{m \times m}$. Let $X_{ij}(t)$ be a \textit{count} of the number of times Player (vertex) $i$ has played strategy $j \in \{1,\dots,m\}$ at time $t$. If:
\begin{equation}
M_i(t) = \sum_{j} X_{ij}(t)
\end{equation}
and:
\begin{equation}
x_{ij}(t) = \frac{X_{ij}(t)}{M_i(t)},
\end{equation}
then the vector $\Vector{x}_i(t) = \langle{x_{i1}(t),\dots,x_{in}(t)}\rangle$ represents the current mixed strategy of Player $i$ at time $t$. For simplicity, we will suppress time in the notation unless needed for the remainder of this paper. Suppose the strategy counts of the players change according to the expected payoff rule :
\begin{equation}
\dot{X}_{ij} = X_{ij} \left(\frac{1}{|N(i)|}\sum_{k \in N(i)} \Vector{e}_j \cdot \Matrix{A}\Vector{x}_k\right)
\label{eqn:CountGrowth}
\end{equation}
Here $(\cdot)$ denotes the standard Euclidean dot product and $N(i)$ denotes the graph-theoretic neighborhood of Player (Vertex) $i$. This approach is precisely the one taken in \cite{PG16} when vertices are treated as species while the strategies at each vertex are treated as sub-species. Unlike \cite{PG16} the exact species proportions (vertex counts) are fixed, making the analysis of \cref{eqn:CountGrowth} simpler. For completeness, we note in the dynamics of \cref{eqn:CountGrowth},  it is possible for counts to decrease if 
\begin{displaymath}
\sum_{k \in N(i)} \Vector{e}_j \cdot \Matrix{A}\Vector{x}_k < 0
\end{displaymath}
In this case, we might assume a player ``forgets'' his prior plays. In general, this can be ignored by rescaling $\Matrix{A}$ so it is always positive; additionally we will only be concerned with proportions throughout the remainder of this paper.

Following the derivation in \cite{PG16}  -- i.e. applying the quotient rule to compute $\dot{x}_{ij}$ yields:
\begin{displaymath}
\forall i,j \left\{\dot{x}_{ij} = \frac{1}{|N(i)|} x_{ij}\left(\sum_{k \in N(i)}\left(\Vector{e}_j - \Vector{x}_i \right) \cdot \Matrix{A}\Vector{x}_k\right)\right.
\end{displaymath} 
The constants $|N(i)|$ adjust the flow speed and can be eliminated to obtain the ordinary network replicator:
\begin{displaymath}
\forall i,j \left\{\dot{x}_{ij} = x_{ij}\left(\sum_{k \in N(i)}\left(\Vector{e}_j - \Vector{x}_i \right) \cdot \Matrix{A}\Vector{x}_k\right)\right.,
\end{displaymath} 
which is \cref{eqn:NetReplicator}.

\section{Analysis of the Jacobian of $2\times 2$ Games}\label{sec:AppendixB}
For an arbitrary $2 \times 2$ game, assume the payoff matrix has form:
\begin{displaymath}
\Matrix{A} = \begin{bmatrix} 0 & r\\s & 0\end{bmatrix}.
\end{displaymath}
For the network replicator, (as in the ordinary replicator \cite{Wei95}), an arbitrary payoff matrix can be modified by subtracting or adding (different) constants to each column without changing the structure of fixed points so long as the ordering of the entries remains fixed. Consequently the network replicator for a $2 \times 2$ payoff matrix is 
\begin{displaymath}
\dot{x_i} =
x_i(1-x_i)\left(\sum_{j\in N(i)} r-(r+s)x_j\right).
\end{displaymath} 
Differentiating, we see that the components of the Jacobian matrix $\Matrix{J}(\Vector{x})$ are:
\begin{equation}
\Matrix{J}_{ij}(\Vector{x}) = \begin{cases}
(1 - 2x_i)\left(\sum_{j \in N(i)} r-(r+s)x_j\right) & \text{if $i = j$}\\
-x_i(1-x_i)(r+s) & \text{if $j \in N(i)$}\\
0 & \text{otherwise}
\end{cases}
\label{eqn:Jacobian}
\end{equation}
For fixed point $\Vector{x}^*$, and let $S \subset V$ be the set of vertices that do not have a pure strategy; i.e. if $i \in V$, then $x_i^* \in (0,1)$. Let $G[S]$ denote the subgraph generated by the vertices in $S$. We'll analyze the possible fixed points and eigenvalues of the corresponding Jacobian matrix in cases. 

\textbf{Case I:} If $r$ and $s$ are opposite sign, then:
\begin{displaymath}
\frac{r}{r+s} \not \in [0,1]
\end{displaymath}
and thus there are no vertices with a mixed strategy. In this case $S = \emptyset$ and $G[S]$ has no edges. From \cref{eqn:Jacobian}, the Jacobian matrix must be diagonal with real eigenvalues given by:
\begin{equation}
\lambda_i = (1-2x_i)\left(\sum_{j\sim i} r-(r+s)x_j\right)
\label{eqn:Eigenvalue}
\end{equation}
Consequently $\Vector{x}^*$ is hyperbolic and admits no circulation. Moreover, when $r > 0 > s$, then $r-(r+s)x_j > 0$ for all $j$ because $x_j \in \{0,1\}$. This implies that any eigenvalue $\lambda_i < 0$ if and only if $x_i = 1$. It follows that the only stable equilibrium is the consensus strategy where all players play Strategy 1. Similarly, when $s > 0 > r$, then the only stable equilibrium is the consensus strategy where all players play Strategy 2. This shows that in Prisoner's dilemma type games, the defect strategy is always stable for all players.

\textbf{Case II:} Suppose $r$ and $s$ have the same sign and without loss of generality suppose that $r,s > 0$. In this case, it is possible for $S$ to be non-empty.

From \cref{eqn:Jacobian}, if $j \not \in N(i)$, then $\Matrix{J}_{ij}(\Vector{x}^*) = 0$ for $i \neq j$. So row $i$ of $\Matrix{J}(\Vector{x}^*)$ contains non-zero entries only at the neighbors of $i$. To solve $\det(\Matrix{J}(\Vector{x}^*) - \lambda \Matrix{I}) = 0$, apply row reduction. We have already noted that if $i \in V \setminus S$, then  row $i$ has a single non-zero entry on the diagonal and $\Matrix{J}(\Vector{x}^*)$ has an eigenvalue given by \cref{eqn:Eigenvalue}. If any of these values are positive, then $\Vector{x}^*$ is unstable.

Suppose $i \in S$. By our previous assertion using row reduction on $\Matrix{J}(\Vector{x}^*) - \lambda \Matrix{I}$, we can remove any non-zero element in the columns corresponding to $j \in V \setminus S$, leaving only the rows and columns corresponding to $S$ to be diagonalized. Let $\Adj(G[S])$ be the (symmetric) adjacency matrix of the subgraph $G[S]$. Let $\Matrix{Q}(\Vector{x})$ be the sub-matrix of the partial row-reduction just discussed. For $i \in S$, $\Matrix{J}_{ii}(\Vector{x}^*) = 0$. Careful inspection shows that:
\begin{displaymath}
\Matrix{Q}(\Vector{x}) = -(r+s)\Matrix{D}\cdot \Adj(G[S]) - \lambda \Matrix{I},
\end{displaymath}
where $\Matrix{D}$ is a diagonal matrix with $x_i(1-x_i)$ on the diagonal. Note that $\Matrix{D}$ is positive definite, and thus has a (diagonal) square root, which we denote $\Matrix{B}$. The remaining eigenvalues of the Jacobian are exactly those of of $\Matrix{D}\cdot \Adj(G[S])$. This matrix shares eigenvalues with the symmetric matrix $\Matrix{B} \cdot \Adj(G[S]) \cdot \Matrix{B}$, and thus all these eigenvalues are real by the Principal Axis Theorem. 

Thus we have shown that that there is no circulation in the phase portrait of the network replicator in $2 \times 2$ games because all eigenvalues of the Jacobian matrix must be real. Consequently, any center manifold indicates directions of neutral stability or instability.

If $G[S]$ has any edges, then since $\Tr(\Adj(G[S])) = 0$, it follows that $\Adj(G[S])$ has both a positive and negative eigenvalue. Since $\Matrix{D}$ is positive definite, the positive eigenvalues of $\Adj(G[S])$ imply that $\Matrix{D} \cdot \Adj(G[S])$ has a positive eigenvalue. Similarly, the negative eigenvalues of $\Adj(G[S])$ mean that $\Matrix{D} \cdot \Adj(G[S])$ has a negative eigenvalue. Since $r+s$ can only be zero when $\sgn(r) \neq \sgn(s)$ and we assumed this was not the case, it follows that there is a positive eigenvalue whenever $G[S]$ has an edge. Therefore we have shown that $\Vector{x}^*$ is unstable whenever $G[S]$ has an edge. 

To summarize, we have shown the following two results: 
\begin{enumerate}
\item If $\Vector{x}^*$ is a fixed point and the corresponding subgraph $G[S]$ has an edge, then this fixed point has an unstable manifold. It immediately follows that any interior fixed points are unstable. 

\item For any fixed point $\Vector{x}^*$, of the network replicator with a $2 \times 2$ payoff matrix, the eigenvalues of the Jacobian $\Matrix{J}(\Vector{x}^*)$ are real and therefore for any initial point $\Vector{x}_0 \in \Delta_2^n$, the solution curves will tend to a rest point $\omega(\Vector{x}_0) \in \Delta_2^n$ on the boundary. That is, neither circulation nor chaotic behavior is possible in the network replicator with a $2 \times 2$ payoff matrix. 
\end{enumerate}

In network terms, these results imply that like pure strategies will tend to be adjacent (when possible) in coordination games, while in anti-coordination games, opposite pure strategies will tend to be adjacent, when possible. The latter is illustrated in the main text.

\section{Fixed Points of RPS on $K_3$}\label{sec:AppendixC}
Let
\begin{displaymath}
\mathbf{A} = \begin{bmatrix}0 & -1 & 1 \\ 1 & 0 & -1\\ -1 & 1 & 0\end{bmatrix}
\end{displaymath}
and consider the network replicator on $K_3$. Algebraic analysis shows that the system has an infinite collection of fixed points that can be organized into three classes as shown in \cref{tab:FixedPoints}. The parameters $a, b, c$ and $r, s, t$ used in specifying the boundary fixed points are chosen from the set $\{1,2,3\}$ with elimination. For example, one of the 36 fixed points sets $r = 1$, $s = 2$ and $t = 3$ and $a = 2$, $b = 3$ and $c = 1$ to obtain the fixed point: $x_{1,1} = 0$, $x_{1,2} = p$, $x_{1,3} = 1-p$, $x_{2,1} = 0$, $x_{2,2} = \tfrac{1}{3}(2-3p)$, $x_{2,3} = \tfrac{1}{3}(1+3p)$ and $x_{3,1} = \tfrac{2}{3}$, $x_{3,2} = 0$, $x_{3,3} = \tfrac{1}{3}$ for $p \in \left[0,\tfrac{2}{3}\right]$.
\begin{table}[htbp]
\centering
\begin{tabular}{|l|c|c|}
\hline
\textbf{Strategy Type} & \textbf{Fixed Points}\\
\hline
Pure Strategy & \begin{minipage}{1.5in}\raggedright
\vspace*{0.25em}
$\Vector{x}_1 = \Vector{e}_{i_1}$, $\Vector{x}_2 = \Vector{e}_{i_2}$, $\Vector{x}_3 = \Vector{e}_{i_3}$
\end{minipage}\\
\hline
Boundary & \begin{minipage}{1.75in}\raggedright
$x_{ra} = p$, $x_{rb} = 1-p$, $x_{rc} = 0$\\
$x_{sa} = \tfrac{1}{3}(2-3p)$, $x_{sb} = \tfrac{1}{3}(1 + 3p)$, $x_{sc} = 0$\\
$x_{ta} = 0$, $x_{tb} = \tfrac{1}{3}$, $x_{tc} = \tfrac{2}{3}$\\
\end{minipage}\\
\hline
Interior & $\mathbf{x}_1 = \mathbf{x}_2 = \mathbf{x}_3 = \left\langle{\tfrac{1}{3},\tfrac{1}{3},\tfrac{1}{3}}\right\rangle$\\
\hline
\end{tabular}
\caption{The three classes of fixed points in the rock-paper-scissors replicator dynamic produce an infinite set of possible fixed points.}
\label{tab:FixedPoints}
\end{table}
We can analyze the stability of the fixed points using a reduced dimensional representation by eliminating the redundant equation and variables; i.e., letting $x_{i,3}=1-x_{i,1} - x_{i,2}$ for $i\in \{1,2,3\}$. 

The set of eigenvalues of the Jacobian matrix varies slightly depending on the pure strategy type (e.g., whether the pure strategy contain a representative rock, paper and scissors). Ignoring multiplicities, the possible sets of eigenvalues are:
\begin{displaymath}
\Lambda_{\text{pure}} \in \left\{ \left\{\pm 2\right\}, \left\{-4, \pm 2, 1\right\}, 
\left\{4, \pm 2, -1\right\}, \left\{\pm 1\right\}\right\}
\end{displaymath}
Thus, the pure strategies are hyperbolic with a non-empty unstable manifold. The eigenvalues of the Jacobian matrix about the fixed points on the boundary fall into two classes. and (ignoring multiplicities):
\begin{displaymath}
\Lambda_{\text{boundary}} \in 
\left\{
\left\{-2, 0, 3p, 2-3p, \pm\frac{2}{3} \sqrt{\sigma}\right\},\right.\\
\left.\left\{2, 0, -3p, -2+3p, \pm\frac{2}{3} \sqrt{\sigma}\right\}
\right\}
\end{displaymath}
where $\sigma = 9 p^2-6p-1$. Since $p \in \left[0,\tfrac{2}{3}\right]$, $\sigma \leq 0$, and therefore, these fixed points have stable and unstable manifolds as well (possibly) as slow and center manifolds because $\sqrt{\sigma}$ is pure imaginary. After discussing the interior fixed point, we show a that this system has a special property that allows us to avoid complicated analysis in this case.

The eigenvalues of Jacobian matrix of the interior fixed point with multiplicities are:
\begin{displaymath}
\Lambda_\text{int} = \left\{\pm\frac{2i}{\sqrt{3}},\pm\frac{i}{\sqrt{3}}\right\}.
\end{displaymath}
As we show in the main text, this must be an elliptic fixed point because the divergence of the phase flow is zero everywhere. This also allows us to conclude that the boundary fixed points are non-attracting (i.e., hyperbolic).

As a consequence of volume preservation on the interior of the state space, the following quantity is also conserved in the network replicator with RPS on $K_3$:
\begin{displaymath}
\tau = \prod_{i=1}^3\prod_{j=1}^3 x_{ij}.
\end{displaymath}
This is a novel extension of conservation of strategy products observed in \cite{AL84}. It is also a variation on the result in \cite{EA83} which argues that the replicator on two species preserves a certain volume form; in our case, the volume form is the classical Euclidean volume, consistent with the form of $\tau$.

\section{Sensitive Dependence on Initial Conditions}\label{sec:AppendixE}
To measure the sensitive dependence on initial conditions, we computed the entropy of symbolized trajectories with varying initial conditions for the strategy at Vertex $1$.  The strategies of the other two vertices where initialized at the interior fixed point $\left\langle{\tfrac{1}{3},\tfrac{1}{3},\tfrac{1}{3}}\right\rangle$. To symbolize, space was broken into $\tfrac{1}{10}\times\tfrac{1}{10}$ grids. Then a path $\gamma = (x_{11}[t],x_{12}[t],x_{13}[t])$ for $t \in [0,1000]$ was converted into the corresponding sequence of grids. The ratio of the entropy of this sequence to the possible maximum entropy (of a uniform random variable) was then computed. The results are shown in Fig. 4b of the main text using a temperature scale.  When orbits are started close to the interior fixed point, they remains close to that fixed point and consequently have lower entropy. As the initial condition of Vertex 1 is moved closer to the boundary, the orbit becomes more chaotic and the entropy approaches that of a uniform random variable. Close observation of the figure shows color striation indicative of nested behavior boundaries, as would be expected. 

\begin{figure}[htbp]
\centering
\includegraphics[width=0.6\textwidth]{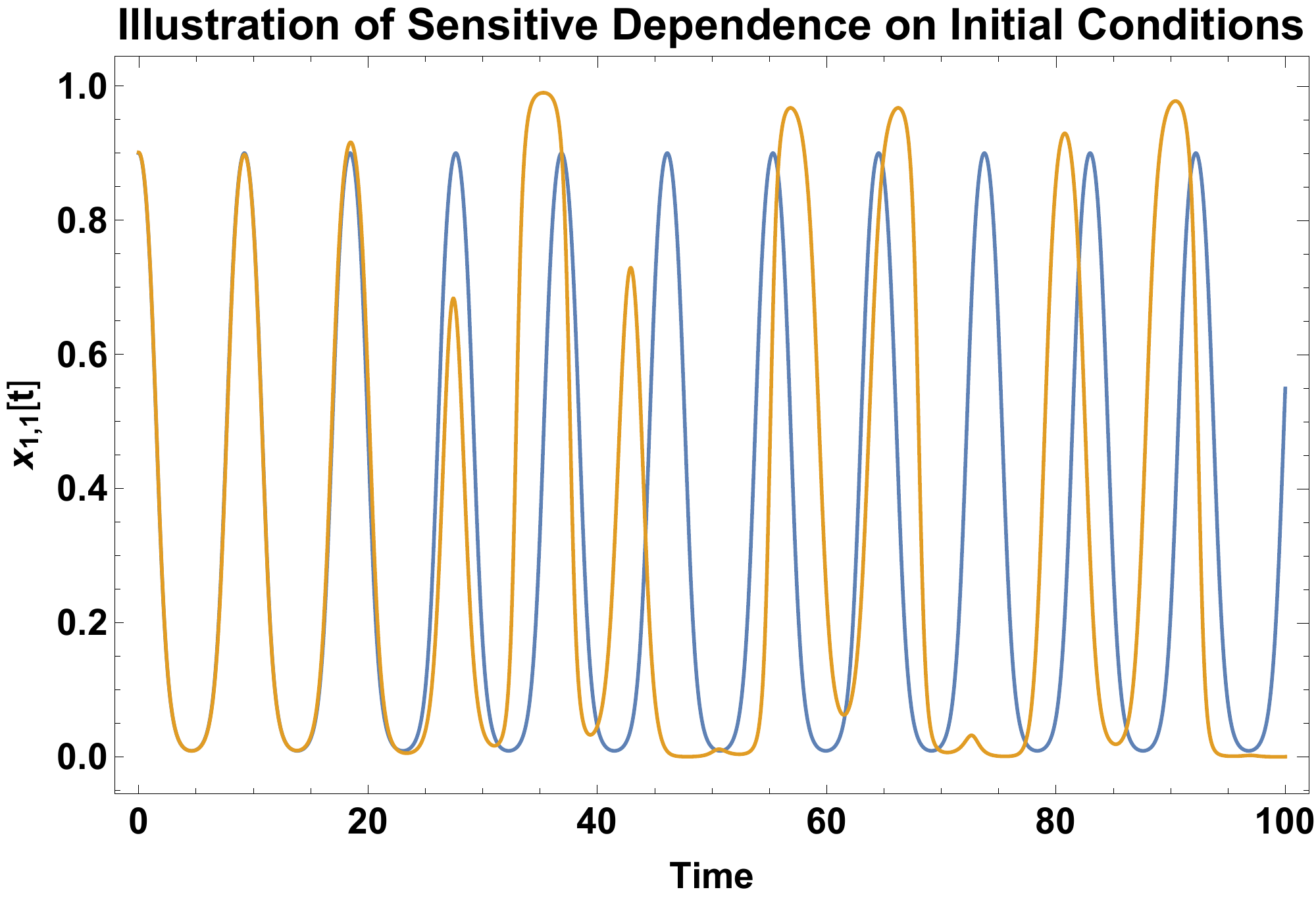}
\caption{Sensitive dependence on initial conditions is illustrated for the network replicator with RPS on $K_3$.}
\label{fig:Dependence}
\end{figure}

To see this effect in specific, \cref{fig:Dependence} shows $\Vector{x}_{1,1}(t)$ when started from two nearby starting points:
\begin{align*}
&\Vector{x}_1(0) = \Vector{x}_2(0) = \Vector{x}_3(0) = \left\langle{\tfrac{9}{10},\tfrac{5}{100},\tfrac{5}{100}}\right\rangle\\
&\Vector{x}_1'(0) = \left\langle{\tfrac{901}{1000},\tfrac{495}{1000},\tfrac{495}{1000}}\right\rangle,\\ 
&\Vector{x}_2'(0) = \Vector{x}_3'(0) = \left\langle{\tfrac{9}{10},\tfrac{5}{100},\tfrac{5}{100}}\right\rangle
\end{align*}
As we expect from a chaotic system, the solutions start close to each other, but after $t = 150$, the dynamics begin to diverge substantially. 

\subsection{Transition to Chaos in Solution Spectra}\label{sec:AppendixF}
The transition from simple (quasi) periodic motion near the fixed point to chaotic motion close to the boundary can be illustrated by an analysis of the spectra of one of the solution components. 
\begin{figure}[htbp]
\centering
\includegraphics[width=0.7\columnwidth]{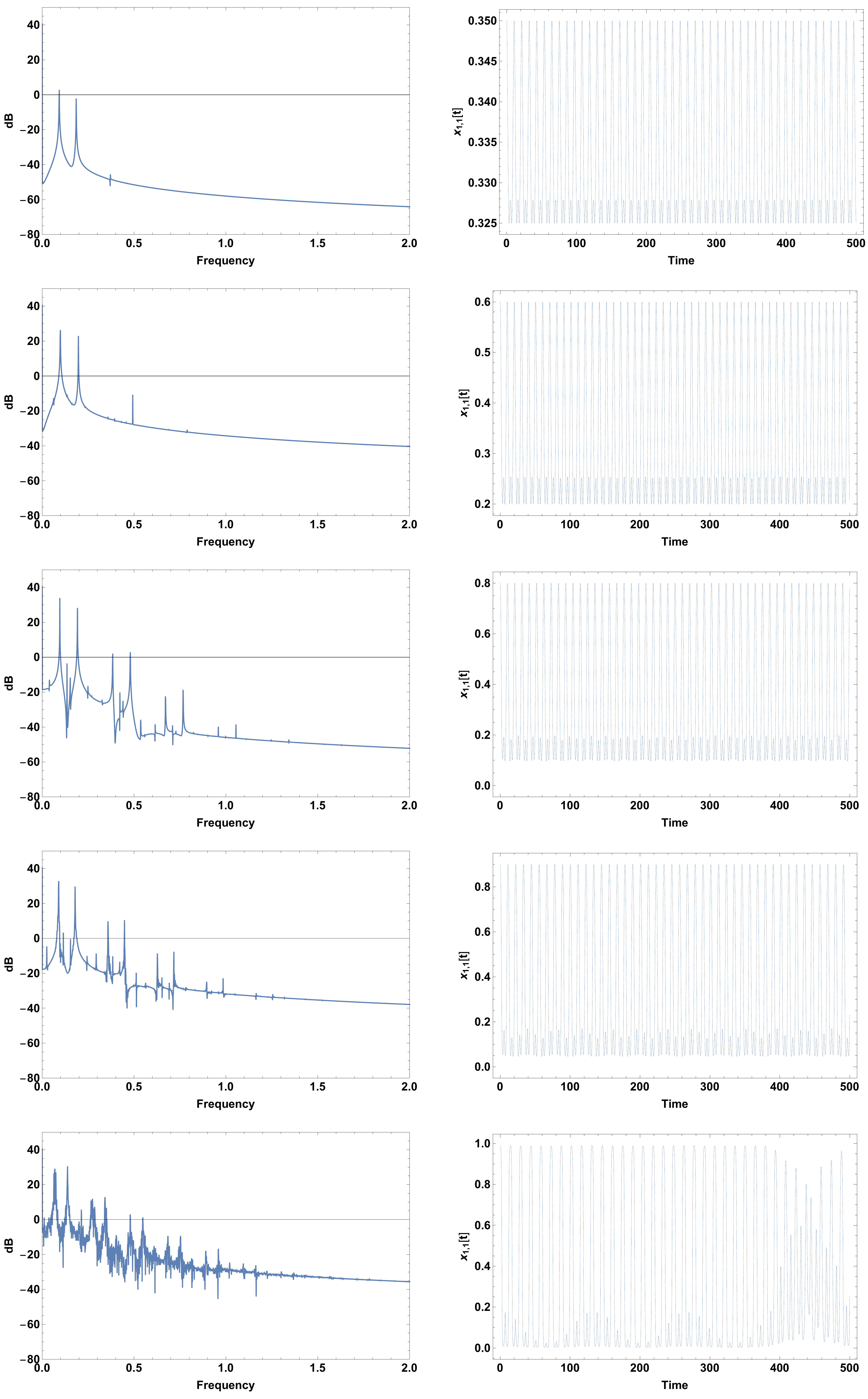}
\caption{The figure illustrates the transition from periodic solutions close to the interior fixed point to a chaotic solution far from the fixed point. In this figure $\Vector{x}_2(0)=\Vector{x}_3(0)=\left\langle{\tfrac{1}{3},\tfrac{1}{3},\tfrac{1}{3}}\right\rangle$. While $\Vector{x}_1(0)$ is constructed so that $x_{12}(0)=x_{13}(0)=\tfrac{1}{2}(1-x_{11}(0))$ always and $x_{11}(0)$ is chosen in the set $\{0.35, 0.6, 0.8, 0.9, 0.99\}$.}
\label{fig:Spectra}
\end{figure}
In \cref{fig:Spectra} the spectrum of $x_{11}(t)$ is computed using a sampling rate of 100Hz. The initial condition of the dynamical system is constructed so that:
\begin{displaymath}
    \Vector{x}_2(0)=\Vector{x}_3(0)=\left\langle{\tfrac{1}{3},\tfrac{1}{3},\tfrac{1}{3}}\right\rangle,
\end{displaymath}
while $\Vector{x}_1(0)$ is constructed so that:
\begin{displaymath}
    x_{12}(0)=x_{13}(0)=\frac{1}{2}\left(1-x_{11}(0)\right)
\end{displaymath} 
and $x_{11}(0)$ is chosen in the set $\{0.35, 0.6, 0.8, 0.9, 0.99\}$. Near the fixed point the spectrum shows two dominant frequencies and the orbit is periodic. As $x_{11}(0)$ increase (toward the boundary), additional frequency components enter the signal. The periodic signal becomes quasi-periodic as the orbit traces out a high-dimensional surface. Interestingly, the signal continues to exhibit these wild swings back toward its initial value. However, this behavior changes after $t=400$ when $x_{11}(0)=0.99$. In this case, a new behavioral regime is entered. We note that the spectrum at this point is rich with frequencies and is consistent with the spectra of other chaotic signals (see e.g., Page 60-61 of \cite{BG96}). 

\section{Neutrally Stable Orbits}\label{sec:AppendixD}
\begin{figure}[htbp]
\centering
\includegraphics[width=0.4\columnwidth]{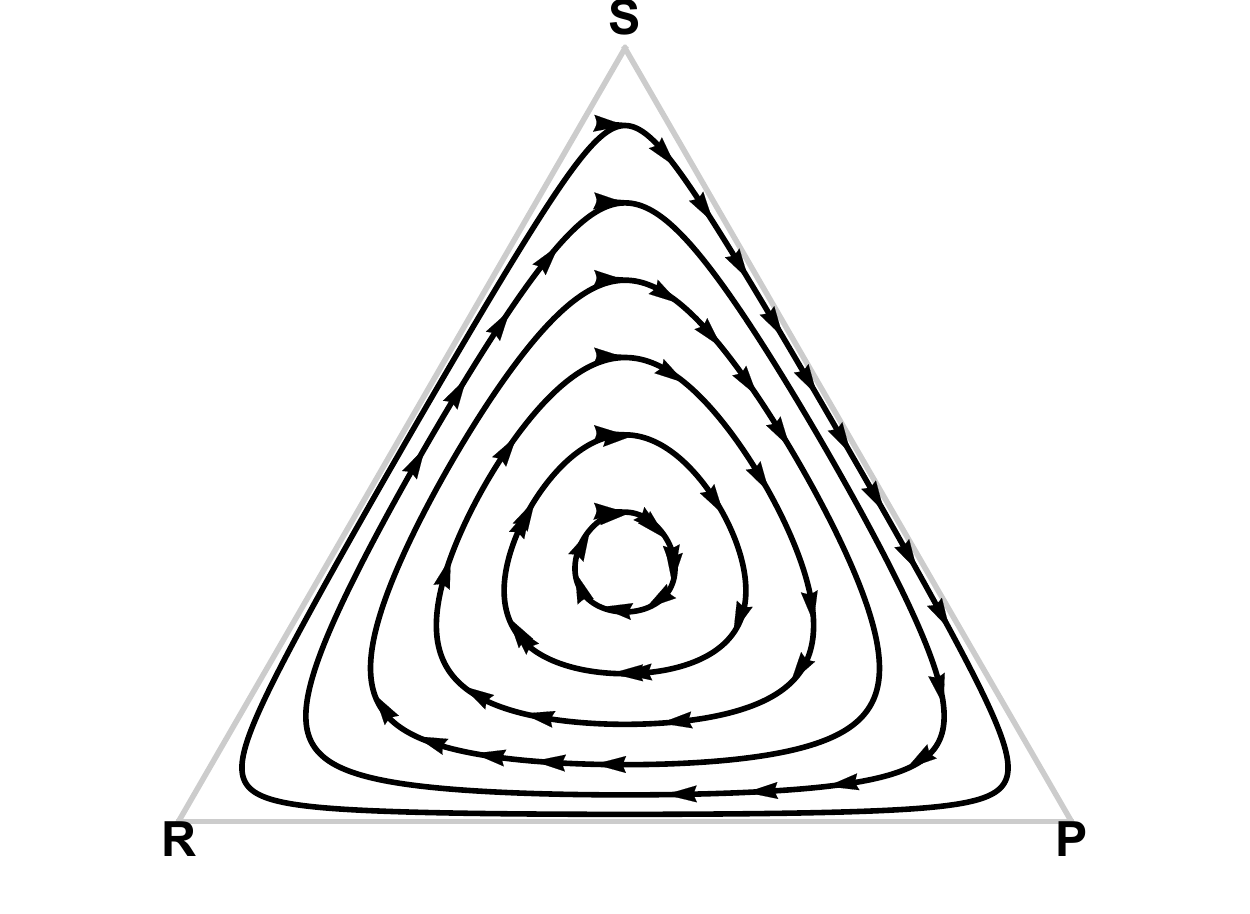}
\caption{A set of neutrally stable orbits exists within the chaotic dynamics of RPS in the network replicator on $K_3$. These orbits act like ordinary RPS in the replicator running backwards in time.}
\label{fig:Backward}
\end{figure}

\begin{figure}[htbp]
\centering
\includegraphics[width=0.75\columnwidth]{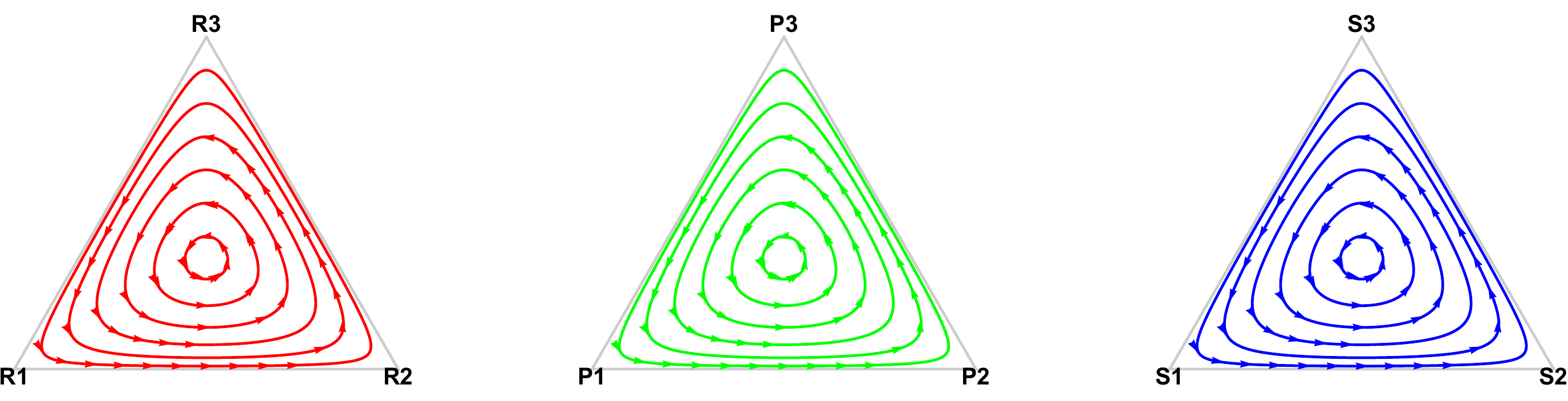}
\caption{Phase portraits for the ternary transform of $(x_{1,k},x_{2,k},x_{3,k})$ illustrating spatial ``chasing'' around the graph.}
\label{fig:Chase}
\end{figure}

Within the dynamics, one can identify neutrally stable cycles that start arbitrarily far from the interior fixed point as well. Simply requiring $x_{1,r}(t) = x_{2,r}(t) = x_{3,r}(t)$ for $r=1,2,3$, the resulting dynamical system has solution curves identical to those of simple ordinary RPS with the replicator dynamic. However, these are not the only neutrally stable cycles that can emerge. Within the chaotic dynamics of the system, there are neutrally stable closed orbits that are identical to the orbits of traditional rock-paper-scissors running backwards in time. To see this, note that if we impose the restriction:
\begin{align}
x_{1,1}(t) = x_{2,2}(t) &= x_{3,3}(t) \label{eqn:Restriction1}\\
x_{1,2}(t) = x_{2,3}(t) &= x_{3,1}(t) \label{eqn:Restriction2}\\
x_{1,3}(t) = x_{2,1}(t) &= x_{3,2}(t) \label{eqn:Restriction3}
\end{align}
then the system of nine differential equations in the network replicator collapses to a system of three differential:
\begin{align}
\dot{x}_{1,1} &= x_{1,1}\left(x_{1,2} - x_{1,3}\right)\label{eqn:iRPS1}\\
\dot{x}_{1,2} &= x_{1,2}\left(-x_{1,1} + x_{1,3}\right)\label{eqn:iRPS2}\\
\dot{x}_{1,3} &= x_{1,3}\left(x_{1,1} - x_{1,2}\right)\label{eqn:iRPS3}
\end{align}
Since the strategies are in order of rock, paper, scissors, these dynamics are precisely the negative of the evolutionary ordinary RPS replicator dynamics; i.e., there are solution curves in this system that cause the ordinary RPS dynamics to run backwards in time for each vertex. Any initial condition satisfying \crefrange{eqn:Restriction1}{eqn:Restriction3} will lead to such curves. This is shown in \cref{fig:Backward}.
This behavior was somewhat surprising, since it runs counter to the ordinary expectation that rock will promote its predator paper, which in turn will promote scissors. The phenomenon can be explained by noting that the populations at the vertices are not self-interacting. Therefore, when \crefrange{eqn:Restriction1}{eqn:Restriction3} hold, then (e.g.) the population of scissors must be growing at the vertex dominated by rock that is adjacent to the vertex dominated by paper. Thus the observed behavior at each vertex will operate in reverse from the ordinary RPS \cite{HS03}. However, spatially, the strategies will move around $K_3$ in a manner consistent with classical RPS. To see this, re-write \crefrange{eqn:iRPS1}{eqn:iRPS3} using \crefrange{eqn:Restriction1}{eqn:Restriction3} to obtain:
\begin{align}
\dot{x}_{1,k} &= x_{1,1}\left(-x_{2,k} + x_{3,k}\right)\\
\dot{x}_{2,k} &= x_{2,1}\left(-x_{3,k} + x_{1,k}\right)\\
\dot{x}_{3,k} &= x_{3,1}\left(-x_{1,k} + x_{2,k} \right)
\end{align}
for $k=1,2,3$. These are the ordinary RPS equations when Vertex 1 acts as rock, Vertex 2 acts as paper and Vertex 3 acts as scissors. The phase portraits for the strategies are shown in \cref{fig:Chase} showing the strategies cycling among the vertices of $K_3$ and cycling in the opposite direction of the trajectories in \cref{fig:Backward}.

\section{Detailed Derivation of the Generalized Hamiltonian}\label{sec:AppendixG}
We show that the general system is not a simple Hamiltonian system, but a generalized Hamiltonian system \cite{P90}. (We note this expression is similar to Eq. (6) of \cite{N73}), which also satisfies Liouville's Theorem.) To see this, we apply the analysis in \cite{Hof96} in which strategy 3 (scissors) is divided out, leaving (again) a 6 dimensional system. In full generality, suppose we have an $m$ strategy game on a graph $G = (V,E)$ with $|V| = n$. Define:
\begin{displaymath}
y_{i,j} = \frac{x_{i,j}}{x_{i,m}} \qquad i\in\{1,\dots,n\},\;j\in\{1,\dots,m-1\}
\end{displaymath}
then by substitution and the quotient rule we have:
\begin{equation}
\dot{y}_{i,j} = \frac{x_{i,j}}{x_{i,m}}\left(\sum_{k \in N(i)} \left(\Vector{e}_j - \Vector{e}_m\right)\cdot\Matrix{A}\Vector{x}_k\right) = \\
\frac{x_{i,j}}{x_{i,m}}\left(\sum_{k \in N(i)} x_{k,m}\left(\Vector{e}_j - \Vector{e}_m\right)\cdot\Matrix{A}\frac{\Vector{x}_k}{x_{k,m}}\right).
\label{eqn:yij}
\end{equation}
Necessarily, $y_{i,m} = 1$ and therefore $\dot{y}_{i,m} = 0$, which we will henceforth ignore. Note that:
\begin{equation}
\frac{1}{x_{k,m}} = 1 + \sum_{l \neq m} \frac{x_{k,l}}{x_{k,m}}
\label{eqn:xkninv}
\end{equation}
because $x_{k,1} + \cdots + x_{k,m} = 1$. Substituting \cref{eqn:xkninv} into \cref{eqn:yij} and noting that $y_{k,j} = x_{k,j}/x_{k,m}$ everywhere yields:
\begin{equation}
\dot{y}_{i,j} = y_{i,j}\left(\sum_{k \in N(i)}\frac{\left(\Vector{e}_j - \Vector{e}_m\right)\cdot\Matrix{A}\Vector{y}_k}{1+\sum_{l\neq m} y_{k,l}} \right)
\end{equation}
Making the substitution:
\begin{displaymath}
u_{i,j} = \log(y_{i,j})
\end{displaymath}
and noting that:
\begin{displaymath}
\dot{u}_{i,j} = \frac{\dot{y}_{i,j}}{y_{i,j}},
\end{displaymath}
we see that:
\begin{displaymath}
\dot{u}_{i,j} = \sum_{k \in N(i)} \frac{\left(\Vector{e}_j - \Vector{e}_m\right)\cdot\Matrix{A}\exp(\Vector{u}_k)}{1+\sum_{l\neq m} \exp(u_{k,l})},
\end{displaymath}
which is \cref{eqn:InteriorU}.

We now construct equations explicitly for RPS. We have:
\begin{displaymath}
\exp(\mathbf{u}_k) = \begin{bmatrix}
e^{u_{k,1}}\\
e^{u_{k,2}}\\
1
\end{bmatrix}
\end{displaymath}
because $y_{k,3} \equiv 1$ and $u_{k,3} = \log(y_{k,3}) = 0$. Substituting this into \cref{eqn:InteriorU} yields:
\begin{align}
\dot{u}_{i,1} &= \sum_{k\in N(i)}\frac{e^{u_{k,1}} - 2e^{u_{k,2}} + 1}{1 + e^{u_{k,1}} + e^{u_{k,2}}}\label{eqn:ui1}  \\
\dot{u}_{i,2} &= \sum_{k\in N(i)}\frac{2e^{u_{k,1}}-e^{u_{k,2}} - 1}{1 + e^{u_{k,1}} + e^{u_{k,2}}} 
\label{eqn:ui2} 
\end{align}
In general note that:
\begin{displaymath}
\frac{e^u - 2e^v + 1}{1 + e^u + e^v} = \frac{1 + e^u + e^v}{1 + e^u + e^v} + \frac{-3e^v}{1+e^u+e^v} = \\
1 - \frac{3e^v}{1+e^u+e^v}
\end{displaymath}
and
\begin{displaymath}
\frac{2e^u - e^v - 1}{1 + e^u + e^v} = \frac{3e^u}{1+e^u + e^v} - \frac{1 + e^u + e^v}{1 + e^u + e^v} = \\
-1 + \frac{3e^u}{1+e^u + e^v} 
\end{displaymath}
Applying these identities to  yields:
\begin{align*}
\dot{u}_{i,1} &= \sum_{k \in N(i)} \left(1 - \frac{3e^{u_{k,2}}}{1+e^{u_{k,1} + u_{k,2}}}\right) \\
\dot{u}_{i,2} &= \sum_{k \in N(i)} \left(-1 + \frac{3e^{u_{k,1}}}{1+e^{u_{k,1} + u_{k,2}}}\right)
\end{align*}
as required. It is now straightforward to see that the Hamiltonian given in Eq. 9 of the main text:
\begin{displaymath}
\mathcal{H} = \sum_{i}\sum_{j} u_{i,j} - \sum_{i} 3\log\left(1 + e^{u_{i,1}} + e^{u_{i,2}} \right)
\end{displaymath}
has the property that:
\begin{displaymath}
\forall i \left\{
\begin{aligned}
\dot{u}_{i,1} &= \sum_{k \in N(i)} \frac{\partial\mathcal{H}}{\partial u_{k,2}}\\
\dot{u}_{i,2} &= \sum_{k \in N(i)} -\frac{\partial\mathcal{H}}{\partial u_{k,1}}
\end{aligned}
\right.
\end{displaymath}
as given in Eq. 10 of the main text. Thus the system is a generalized Hamiltonian system obeying Liouville's Theorem.

\section{Linearization of RPS on $K_4$ near the Interior Fixed Point}\label{sec:AppendixH}
We briefly show that as in the case for $K_3$, near the fixed point the network replicator with RPS on $K_4$ behaves as a degenerate Hamiltonian system. First set $x_{i,3} = 1 - x_{i,1} - x_{i,2}$ for all $i$. This reduces the dimension of the dynamical system from 12 to 8. Linearizing about the interior fixed point we see:
\begin{align}
\dot{x}_{i,1} &= -\frac{1}{3}\left(\sum_{k\neq i} x_{k,1} + 2 x_{k,2} \right)\\
\dot{x}_{i,2} &= \frac{1}{3}\left(\sum_{k\neq i} 2x_{k,1} + x_{k,2} \right)
\end{align}
As conjugate momenta, define:
\begin{align}
p_{i,1} &= \sum_{k\neq i} x_{k,2}\\
p_{i,2} &= \sum_{k\neq i} x_{k,1} 
\end{align}
Then we see that:
\begin{align}
\dot{x}_{i,1} &= -\tfrac{1}{3}\left(p_{i,1} + 2p_{i,2}\right)\\
\dot{x}_{i,2} &= \tfrac{1}{3}\left(2p_{i,1} + p_{i,2}\right)\\
\dot{p}_{i,1} &= \tfrac{1}{3}\left(6x_{i,1} + 3x_{i,2} + \sum_{k \neq i} 4x_{j,1} +  \sum_{k \neq i} 2x_{j,2}\right)\\
\dot{p}_{i,2} &= \tfrac{1}{3}\left(3x_{i,1} + 6x_{i,2} + \sum_{k \neq i} 2x_{j,1} +  \sum_{k \neq i} 4x_{j,2}\right)
\end{align}
It is possible to construct an explicit Hamiltonian:
\begin{multline}
\mathcal{H} = \frac{1}{3}\left(
\sum_{i} p_{i,1}^2 + p_{i,2}^2 + p_{i,1}p_{i,2} + 
\sum_{i} 3x_{i,1}^2 + 3x_{i,2}^2 +
\sum_{i} 3x_{i,1}x_{i,2} +\right. \\
\left.
\sum_i\sum_{j>i} 4x_{i,1}x_{j,1} + \sum_i\sum_{j>i} 4x_{i,2}x_{j,2} + \sum_{i}\sum_{j>i} 2x_{i,1}x_{j,2} + 
\sum_{i}\sum_{j>i} 2x_{i,2}x_{j,1}\right)
\end{multline}
However the fact that the time derivatives of the conjugate momenta can be expressed solely in terms of the state variables and the time derivatives of the state variables can be expressed solely in terms of the conjugate momenta is sufficient to show that the system is a Hamiltonian system.

\bibliographystyle{IEEEtran}
\bibliography{NetworkReplicatorChaos}

\end{document}